\documentclass[a4paper,11pt]{article}
\pdfoutput=1 

\usepackage{jcappub} 

\usepackage[T1]{fontenc} 
\usepackage{caption}
\usepackage{subcaption}

\title{\boldmath GravitoMagnetic Force in Modified Newtonian Dynamics}

\author{Qasem Exirifard}

\affiliation{
Institute for Research in Fundamental Sciences (IPM),
\\Tehran, Iran
}

\emailAdd{exir@theory.ipm.ac.ir}

\abstract{We introduce  the Gauge Vector-Tensor (GVT) theory  by extending the AQUAL's approach  to  the GravitoElectroMagnetism (GEM) approximation of gravity.   GVT is a generally covariant theory of gravity  composed of a pseudo Riemannian metric and two $U(1)$ gauge connections that reproduces  MOND in  the limit of very weak gravitational fields  while remains consistent with the Einstein-Hilbert gravity in the limit of strong and Newtonian gravitational fields.   GVT also provides a simple framework to study the GEM approximation to gravity. We illustrate that  the  gravitomagnetic force at the edge of a galaxy can  be in accord with either GVT  or $\Lambda$CDM but not both.   We also study the physics of the GVT theory around  the  gravitational saddle point of the Sun and Jupiter system. We notice that the conclusive  refusal of  the GVT theory demands measuring    either  both of the gravitoelectric and gravitomagnetic fields inside the Sun-Jupiter  MOND window, or the gravitoelectric field inside two different solar GVT MOND windows. The GVT theory, however, will be favored by observing an anomaly in the gravitoelectric field inside a single MOND window.
}

\begin{document}

\maketitle
\flushbottom

\section{Introduction}
Either $95\%$ of the observed universe is made of things that have not yet been observed in the Solar system, or the law of dynamics or gravity should be modified in  very low accelerations or very weak gravitational fields. The  $\Lambda$-CDM model of cosmology buys the first approach. Its challenges \cite{Famaey:2011kh,Kroupa:2013yd}, however,  signal that "the physics of the dark sector is, at the very least, much richer and complex than currently assumed, and that our understanding of gravity and dynamics might also be at play" \cite{Famaey:2013ty}.   The second approach is the modified theories of gravity. The modified theories of gravity can be classified into the following two categories:
\begin{enumerate}
\item Phenomenological search for  the dynamics of the metric.
\item Introducing new degrees of freedom for gravity in addition to the metric.
\end{enumerate}
The first class assumes that gravity is described by a  pseudo Riemannian metric and the action of gravity is given by
\begin{equation}
S = \int \sqrt{-\det g} \big({\cal L}_g[g_{\mu \nu}, R_{\mu\nu\lambda\eta}, \nabla_\mu]+{\cal L}_m[\Psi]\big)\,,
\end{equation}
where $R_{\mu\nu\lambda\eta}$ is the Riemann tensor constructed out of the metric $g_{\mu\nu}$, ${\cal L}_m[\Psi]$ is the matter's action not necessarily minimally coupled to the metric, and ${\cal L}_g$ is the gravity's action. The Einstien-Hilbert theory assumes ${\cal L}_g=R$ where $R$ is the Ricci scalar. The purchasers of   this class choose to reject the  Einstein-Hilbert assumption and search for families of ${\cal L}_g$ reproducing the dynamics of nature in large scales. Considering the infinite number of possibilities in choosing $\cal L$ and the finite set of the cosmological data, this purchase will work \cite{Exirifard:2008dy}. It would  not necessarily be in accord  with the principle of the Occam's razor.  It also will lead to a set of nonlinear partial differential equations of degrees larger than two, a set of equations which most of  rational humans  would  despise. These are, however, the prices to pay. 

The second class of the modified theories of gravity introduces new degrees of freedom in addition to the metric to describe gravity. The most known example of this class is the TeVeS theory  \cite{Bekenstein:2004ne}. TeVeS introduces a pseudo Riemannian metric, a scalar and a vector field in order to phenomenologically describe the physics  in very weak gravitational fields (the MOND regime) . The TeVeS theory defines new nonlinearity in order to solve the physics of the MOND regime. The introduced  nonlinearity, however, is not local to the MOND regime of the theory. It continues to the very strong gravity regime of theory.   The physics of very strong gravitational systems, therefore,  strongly constrain the TeVeS theory. This signals that the introduced nonlinearity of the TeVeS is not appropriate to describe the physics of the MOND regime. One should define a nonlinearity capable of producing the physics of the MOND system such that  the nonlinearity does not propagate all the way down to the Newtonian and strong regime of the theory. In order to  perform such a definition, we go back to  the very root of the TeVeS theory: the AQUAL theory. We show how to apply the AQUAL procedure upon the GravitoElectroMagnetism approximation of gravity. This introduces a non-covariant version for GEM in MOND regime whose generally covariant version demands introducing gauge vector fields rather than a scalar field. We, thus, introduce two $U(1)$ gauge vectors in addition to the metric and present a generally covariant theory for GEM in the MOND regime. This theory, which we call the Gauge Vector Tensor theory, reproduces  MOND in  the limit of very weak gravitational fields  while remains consistent with the Einstein-Hilbert gravity in the limit of strong and Newtonian gravitational fields.   In contrary to the TeVeS theory,  the GVT theory is in total agreement with the physics of the strong gravity. Its equations of motion are also much simpler than those of the TeVeS theory.

The paper is organized as follows: Sections \ref{subsectionGEM1} and \ref{subsectionGEM2}  review the GravitoElectroMagnetism (GEM) to gravity.   Section \ref{reviewAQUALmodel} reviews the algorithm that leads to the AQUAL theory as a realization of the MOND paradigm. Section \ref{upliftAQUAL}  applies the AQUAL's algorithm to  GEM.  Section \ref{section6} introduces one gauge field and presents a covariant realization of the GEM to MOND. It also discusses the phenomenological constraints on the theory. Section \ref{section7} introduces an additional gauge field in order to make the theory fully consistent with the predictions of the Einstein-Hilbert theory for the strong and Newtonian gravitational field. Section \ref{section8} studies various regimes of the GVT theory.  The GVT theory possesses the Newtonian and strong regime of gravity, the MOND regime and the post-MONDian regime. Section \ref{section9} calculates the gravitomagnetic field of a spinning galaxy in the $\Lambda$CDM theory and the GVT theory. It shows that the gravitomagnetic field of a galaxy can be in accord with only one of them.   Section \ref{fasle4-2} studies the physics of the GVT theory around  the  gravitational saddle point of the Sun and Jupiter system. It notices that the conclusive  refusal of  the GVT theory demands measuring    either  both of the gravitoelectric and gravitomagnetic fields inside the Sun-Jupiter  MOND window, or the gravitoelectric field inside two different solar GVT MOND windows. It concludes that the GVT theory, however, can be favored by observing an anomaly in the gravitoelectric field inside a single MOND window. Section \ref{section11} provides the conclusion and outlook.

\section{Response of the test probes to the gravitomagnetic field}
\label{subsectionGEM1}
Classical gravity is governed by a single scalar field, the gravitational potential. The Newtonian gravitational potential satisfies:
\begin{equation}\label{phiNewton}
\nabla^2 \Phi = 4 \pi G \rho,
\end{equation} 
where $\rho$ is the density of matter. Albert Einstein  attempting  to uplift gravity to a relativistic  regime, first replaced the space-time metric of Minkowski by
\begin{equation}
ds^2 \,=\,  - c(\Phi)^2 dt^2 + dx^2 + dy^2+ dz^2,
\end{equation}
later with the Gromann's help, he introduced  the Riemannian metric,
\begin{equation}
ds^2 \,=\,  g_{\mu\nu} dx^\mu dx^\nu,
\end{equation}
 as  the relativistic gravity \cite{Goenner:2008rr}.  The  relativistic theory of gravity has a symmetric rank-two tensor: the metric. The metric has 10 components in  four dimensions, 9 more than the degrees of the classical gravity.  To perceive the physical meaning of the degrees of the freedom of the relativistic theory, let the trajectory of a slow moving particle  be considered  in a  static deviation from  the Minkiowki metric. In so doing, the metric reads 
\begin{eqnarray}
ds^2 &=&  \eta_{\mu\nu} dx^\mu dx^\nu + h_{\mu\nu} dx^\mu dx^\nu\,,\\
h_{\mu\nu} & = & h_{\mu\nu}(x^i)\,, \\
h_{\mu\nu} &\ll& 1\,.
\end{eqnarray}
Only for a relativistic mass distribution like a geon \cite{Geons} the off-diagonal components of  $h_{ij}$  are comparable to its other components.  The contribution of the $h_{ij}$ are also suppressed for the  orbits of slow moving particles. We are considering the geometry around a non-relativistic mass distribution.  We also study the orbits of massive slow moving particles.  In these circumstances the orbit of the particles can be derived from
\begin{equation}
\label{action}
S = -\frac{m}{2} \int d\tau g_{\mu\nu} \dot{x}^\mu \dot{x}^\nu\approx-\frac{m}{2} \int d\tau \left(c^2(-1+2 A_0)\dot{t}^2 + (\dot{x}^i)^2  + 2  c A_{i} \dot{t} \dot{x}^i \right)\,,
\end{equation} 
wherein $\sum_{i\neq j}h_{ij} \dot{x}^i \dot{x}^j $  has been ignored, and $\tau$ is an affine parameter and
\begin{subequations}
\label{moffat2.8}
\begin{eqnarray}
c^2 A_{0} &\equiv & \frac{1}{2} h_{00}\,,\\
c A_{i} & \equiv & h_{i0}\,,
\end{eqnarray}
\end{subequations}
and 
\begin{equation}
{}^. = \frac{\partial}{\partial \tau}\,.
\end{equation}
The Euler-Lagrange equation for $t$ derived from \eqref{action} reads
\begin{equation}\label{eq2.8}
c^2 \dot{t} (-1+ 2 A_0) + 2  c A_i \dot{x}^i \,=\, cte\,,
\end{equation} 
where $\dot{t} (-1+ 2 A_0)$ stands for the gravitational redshift while $2 A_{i} \dot{x}^i$ represents a  relativistic term\footnote{This is the kind of the modification of the effective energy of a particle that lets the  extraction of energy  from a black hole ( the Penrose mechanism) \cite{Penrose}.  }.  Eq. \eqref{eq2.8} can be solved for $t$ in term of $\tau$:
\begin{equation}\label{tau=t}
t = \tau + O(A)\,,
\end{equation}
wherein appropriate unite of time is chosen. 
The Euler-Lagrange equation for $x^i$ derived from \eqref{action} then leads to
\begin{equation}\label{xac0}
\frac{d^2 x^i}{d\tau^2} = c^2 \partial^i A_{0} + c \delta^{ij} (\partial_k A_j - \partial_j A_k) \frac{dx^k}{d\tau}\,.
\end{equation}
Utilizing \eqref{tau=t} then results
\begin{equation}\label{xac}
\frac{d^2 x^i}{dt^2} = c^2 \partial^i A_{0} + c \delta^{ij} (\partial_k A_j - \partial_j A_k) \frac{dx^k}{dt} + O(A^2)\,.
\end{equation}
Now let it be redefined  
\begin{eqnarray}
A_{0} &\to &\frac{1}{c^2} A_0= \Phi\,,\\
A_{i} &\to & \frac{1}{c^2} A_i\,,
\end{eqnarray}
using which the equation \eqref{xac} can be rewritten as follows
\begin{equation}
\label{appendinxneeds}
\ddot{x} = -\nabla \Phi +   \frac{v}{c} \times  (\nabla \times A)\,.
\end{equation}
This allows interpreting  $\nabla \times A$ as a gravitomagnetic field. $\nabla \times A$ causes precessions of the orbits of a test particle. This precession is referred to as  the Lense-Thirring precession \cite{Lense}.  Ref. \cite{Iorio:2010rk} provides a decent recent review on  Lense-Thirring precession for planets and satellites in the Solar system. 
The similarity between the gravitomagnetic field and magnetic field beside the spin precession formula in electrodynamics ($\dot{S}= \mu \times B , \mu = \frac{e}{2m} S$)  dictates that  the spin  of a gyroscope  precesses  by  \cite{padi}
\begin{equation}
\Omega_{LT} = - \frac{1}{2}\nabla \times A\,.
\end{equation}
This precession  is called  the Pugh-Schiff frame-dragging precession \cite{Pugh, Schiff}. The Pugh-Schiff frame-dragging  precession  due to the rotation of the earth recently has been measured by the gravity probe B with the precision of 19\% \cite{Everitt: }. GINGER,  aiming to improve the sensitivity of the ring resonators,   plans  to measure  the gravitomagnetic effect  with a precision at least one order better than that of the gravity probe B \cite{Tartaglia:2012fd}. Also  LAGEOS and LAGEOS 2, and with a number of GRACE (Gravity Recovery and Climate Experiment)  have confirmed the prediction of Einstein General Relativity for the Earth's gravitomagnetic field with with an accuracy of approximately 10\% \cite{LAGEOS}.   Ref. \cite{Murphy:2007nt} shows that   the gravitomagnetic field of the Earth  is in agreement with the Einstein theory's prediction with approximately 0.1\% accuracy via lunar laser ranging (LLR).  

\section{GravitoElectroMagnetism approximation }
\label{subsectionGEM2}
In the linearized Einstein-Hilbert gravity, the Einstein field equations written in the harmonic gauge simplifies  to 
\begin{eqnarray}
\Box \bar{h}_{\mu\nu} &=& -16 \pi G T_{\mu\nu}^{(0)} \,,\\
\label{HarmonicGauge}
\partial_\mu  \bar{h}^{\mu\nu} &=&0\,,
\end{eqnarray}
where $\bar{h}_{ij}$ is the trace reversed perturbation 
\begin{equation}\label{htracereversed3.3}
\bar{h}_{\mu\nu} \, \equiv \, h_{\mu\nu} - \frac{1}{2} \eta_{\mu\nu} h^{\alpha}_{~\alpha} \,.
\end{equation}
We notice that the linearized equations can be derived from 
\begin{equation}\label{4.4action}
S  = \int d^4 x (- \frac{1}{2}\partial_\alpha \bar{h}_{\mu\nu} \partial^\alpha \bar{h}^{\mu\nu} + 16 \pi G \bar{h}_{\mu\nu} T^{\mu\nu} + \lambda_\nu \partial_\mu \bar{h}^{\mu\nu})\,,
\end{equation}
 where $\lambda_\nu$ is a local Lagrange  multiplier enforcing \eqref{HarmonicGauge}, and $T^{\mu\nu}$  represents the linear energy-momentum tensor. Note that the effective action is invariant under the residual symmetry of the harmonic gauge. It is invariant under
\begin{subequations} \label{RHG}
\begin{eqnarray}
 h_{\mu\nu} & \to & h_{\mu\nu} + \partial_\mu \xi_\nu + \partial_\nu \xi_\mu \,,\\
 \Box \xi_\mu & = & 0\,.
\end{eqnarray}  
\end{subequations}
We do not fix the residual symmetry. We consider it as the symmetry of  the action. 
 We decompose $\bar{h}_{\mu\nu}$ to
 \begin{equation}\label{metricA}
 \bar{h}_{\mu\nu} dx^\mu dx^\nu \equiv  \bar{A}_0 dt^2 + \bar{A}_i dx^i dt + \bar{h}_{ij} dx^{i} dx^{j}\,.
 \end{equation}
 Inserting \eqref{metricA} into \eqref{4.4action} yields:
 \begin{equation}\label{4.7action}
S  = \int d^4 x (\frac{1}{2} \bar{A}_0 \Box \bar{A}_0 + \bar{A}_i \Box \bar{A}_i +\frac{1}{2} \bar{h}_{ij} \Box \bar{h}^{ij} + 16 \pi G \rho \bar{A}_0 + 32 \pi G \bar{A}_i J^i + 16 \pi G \bar{h}_{ij} T^{ij})\,,
\end{equation}
and the constraints read
\begin{eqnarray}\label{Lorentz}
\partial^\mu \bar{A}_\mu & =&0\,, \\
\label{wave}
\partial_0 \bar{A}_i - \partial_j  \bar{h}^{j}_{~i} &=& 0 \,.
\end{eqnarray}
The action of  \eqref{4.7action} at the level of the equations of motion is equivalent to
\begin{equation}\label{4.10action}
S  = \int d^4 x (-\frac{1}{2}\bar{A}_0 \Box \bar{A}_0 + \frac{1}{2}\bar{A}_i \Box \bar{A}_i + \frac{1}{2}\bar{h}_{ij} \Box \bar{h}^{ij} +  16 \pi G \bar{A}_i J^i -16 \pi G \rho \bar{A}_0  +16 \pi G \bar{h}_{ij} T^{ij})\,.
\end{equation}
Now let it be defined:
\begin{eqnarray}
\bar{A}_\mu & \equiv &(\bar{A}_0, \bar{A}_i)\,,\\
J_\mu & \equiv & (\rho, J_i)=T_{0\mu}\,.
\end{eqnarray}
Then \eqref{4.10action}  simplifies  to 
\begin{equation}\label{4.13action}
S  = \int d^4 x (\frac{1}{2} \bar{A}_\mu \Box \bar{A}^\mu+ \frac{1}{2} \bar{h}_{ij} \Box \bar{h}^{ij} +  16 \pi G \bar{A}_\mu J^\mu +16 \pi G \bar{h}_{ij} T^{ij})\,,
\end{equation}
Eq. \eqref{Lorentz} yields: 
\begin{equation}\label{4.14AAFF}
\frac{1}{2} \int d^4 x \bar{A}_\mu \Box \bar{A}^\mu = \frac{1}{2} \int d^4x \bar{A}_\mu (\Box \eta^{\mu\nu} -\partial^\mu \partial^\nu) \bar{A}_\nu = -\int d^4x  \frac{1}{4} \bar{F}_{\mu\nu} \bar{F}^{\mu\nu}\,,
\end{equation}
where
\begin{equation}
\bar{F}_{\mu\nu}  \equiv  \partial_\mu \bar{A}_\nu -\partial_\nu \bar{A}_\mu\,\,.
\end{equation}
Utilizing \eqref{4.14AAFF} re-expresses  \eqref{4.13action} to: 
\begin{subequations}
\label{SA}
\begin{eqnarray}
S & = & \int d^4 x ({\cal L}_g + {\cal L}_S + {\cal L}_\lambda)\,, \\
{\cal L}_g & \equiv & -\frac{1}{4} \bar{F}_{\mu \nu} \bar{F}^{\mu\nu} - \frac{1}{2} \partial_\alpha \bar{h}_{ij} \partial^\alpha \bar{h}^{ij} \,, \label{FHaction}\\
{\cal L}_s & \equiv& 16 \pi G \bar{A}_{\mu} J^\mu + 16 \pi G \bar{h}_{ij} T^{ij} \,,\label{source}\\
{\cal L}_\lambda & \equiv & \lambda_0 \partial^\mu \bar{A}_\mu + \lambda^i (\partial_0 \bar{A}_i - \partial_j \bar{h}_{i}^{~j})\,,\label{gaugefixing}
\end{eqnarray}
\end{subequations}
The first term of ${\cal L}_g$ in \eqref{FHaction} is  the GravitoElectoMagnetic (GEM) approximation to gravity.  ${\cal L}_s$ in \eqref{source} describes how the fields couple to the sources (Energy momentum tensor). ${\cal L}_\lambda$ in \eqref{gaugefixing} is the gauge fixing Lagrangian. In comparison to the electrodynamics, the equations of motion for $\lambda$  impose two extra conditions of \eqref{Lorentz} and \eqref{wave} on $\bar{A}_\mu$.
Eq.  \eqref{Lorentz}  states that the GEM should be solved in the Lorentz gauge.   Eq. \eqref{wave} implies that GEM has wave solutions only if  $\bar{h}_{ij}$ field possesses a wave solution.  The wave solution  is due to the dynamics of the $\bar{h}_{ij}$ field. This means that though the GEM is akin to the ordinary electrodynamics it lacks radiation.

Near and around the galaxies, $\bar{h}_{ij}$ is suppressed due to the non-relativistic velocity of the stars and gas inside the galaxy.  At the leading order $\bar{h}_{ij}$ also does not affect the orbits of slow-moving massive particles. Slow moving particles see only the GEM part of the metric \eqref{action}. Since we are interested in the orbits of  slow moving massive particles around a galaxy we just consider only the GEM part of  \eqref{SA}:
\begin{subequations}
\label{SA2}
\begin{eqnarray}
S & = & \int d^4 x ({\cal L}_A + 16 \pi G \bar{A}_\mu J^\mu +  \lambda   \partial^\mu \bar{A}_\mu )\,,\\\label{SA2L}
{\cal L}_A & = & -\frac{1}{4} \bar{F}_{\mu \nu} \bar{F}^{\mu\nu} \,.
\end{eqnarray}
\end{subequations}
 Also note that  time dependent $A_\mu$ , through the constraint equation \eqref{wave}, induces a time-dependent  behavior for  $h_{ij}$. The orbits of the stars at the leading approximation are blind to the change in $h_{ij}$. In the study of the orbits of the stars, therefore, the time dependent solutions of \eqref{SA2} are valid. 
The symmetry of the truncated Lagrangian \eqref{SA2L} is 
\begin{equation}\label{U1symmetry}
\bar{A}_{\mu} \to \bar{A}_{\mu} + \partial_\mu \Lambda\,,
\end{equation}
where $\Lambda$ is a general scalar field.    Part of this symmetry is broken by the gauge fixing Lagrangian. 

\section{AQUAL  as a Realization of MOND}
\label{reviewAQUALmodel}
The Newtonian approximation of the linearized GEM action \eqref{SA2} reads
\begin{eqnarray}
\label{moffat4.1}
\bar{A}_\mu(x,t) &= & (4\Phi(x),\vec{0})\,,\\
J_{\mu} & = & (\rho(x), \vec{0}) \,,
\end{eqnarray}
Notice that $\bar{A}_0$ is equal to $4\Phi$ rather than $2\Phi$ because  $\bar{A}_\mu$ comes from the trace reversed metric. For the reversed trace metric  $\bar{h}_{00}=4  \phi$ and $\bar{h}_{ij}=0$  give  $h_{00}=2 \phi$ and $h_{ii}=- 2\phi$. 
Inserting the Newtonian approximation into \eqref{SA2} yields
\begin{eqnarray}\label{4.3factor}
S & = &-16  \int d^4 x\, (\frac{1}{2} |\nabla\phi|^2  +4  \pi G \rho \phi) \,.
\end{eqnarray}
Notice that \eqref{4.3factor}  up to the  overall factor of  
\begin{equation} \label{overalfactor}
16\int dt\,,
\end{equation}
is equivalent to the Newtonian gravitational action:
\begin{equation}
S_{N} = -\int d^3x \left[|\nabla \Phi|^2 + 8\pi G \rho \Phi\right]\,.
\end{equation}
In the Modified gravity realization of the MOND \cite{MOND}, one replaces the Newtonian classical field theory  with a general field theory but retain the Newtonian dynamics ($F= m a $):
\begin{equation}\label{SMoG0}
S_{MoG} =- \int d^3x \left[{\cal L}(\Phi, \nabla \Phi,\cdots) + 8\pi G \rho \Phi\right]\,.
\end{equation}
Keeping intact the Newtonian dynamics means that the orbits of slow moving particles are derived from  \eqref{action}.
The AQUAL approach  \cite{AQUAL}  assumes that the symmetries for the equations of motions derived from $S$ and $S_{MoG}$ are the same. The symmetries for  $S$ are
\begin{eqnarray}\label{phisymmetry}
\Phi \to \Phi +  {\Lambda}\,,
\end{eqnarray}
where  ${\Lambda}$  is constant. Imposing \eqref{phisymmetry} on \eqref{SMoG0} requires $\cal L$ to be a functional of the derivative of the Newtonian potential:
\begin{equation}
{\cal L}\,\equiv \,{\cal L}(\nabla \Phi, \nabla^2 \Phi,\cdots) \,.
\end{equation} 
AQUAL also requires the equations to be second order. So the Lagrangian is simplified to 
\begin{equation}
{\cal L}\,\equiv \,{\cal L}(\nabla \Phi) \,.
\end{equation} 
We can construct only one scalar out of $\nabla \Phi$. So the Lagrangian reads
\begin{equation}
{\cal L}\,\equiv \, {\cal F}(\frac{1}{a_0^2}\nabla \Phi.\nabla \Phi) \,,
\end{equation} 
and the AQUAL action follows
\begin{equation}\label{SMoG}
S_{AQUAL} =- \int d^3x \left[a_0^2 {\cal F}( \nabla \Phi.\nabla \Phi/a_0^2) + 8\pi G \rho \Phi\right]\,.
\end{equation}
The first variation of the AQUAL action with respect to $\Phi$ yields
\begin{equation}\label{phiMOND0}
\nabla^\alpha\left( \mu(\frac{|\nabla \Phi|}{a_0}) ~\nabla_\alpha \Phi \right)= 4 \pi G \rho,
\end{equation} 
where
 \begin{equation}\label{muF}
 \mu(x) \,\equiv \, \frac{d{\cal F}(y)}{dy}|_{y=x^2}= {\cal F}'(x^2).
 \end{equation}
 The MOND terminology than requires \cite{MONDa0}\footnote{The most widely used from $\mu$ are \cite{AQUAL,mu}: $\mu(x)=\frac{x}{x+1}$ and $\mu(x) = \frac{x}{\sqrt{1+x^2}}$.}
\begin{subequations}
\label{moffat413}
\begin{eqnarray}\label{HEregime}
\mu(x)\approx 1 ~:~  \text{For} ~x \gg1\,,\\ \label{DeepMOND} 
\mu(x) \approx x ~:~ \text{For} ~x \leq 1\,,
\end{eqnarray}
\end{subequations}
and
\begin{equation}
\label{valuea0mo}
a_0 = (1.0 \pm 0.2 ) \times 10^{-10} \frac{m}{s^2}\,.
\end{equation}

\section{AQUAL  Extension  to  GEM}
\label{upliftAQUAL}
Following the AQUAL model, we search for a non-linear generalization of \eqref{SA2} that leads to  second-order differential equations. This generalization must coincide  to  the AQUAL model for a vanishing gravitomagnetic field. We are assuming that the physics of the MOND regime follows from $\bar{h}_{ij}=0$ and $\bar{h}_{00}\neq 0$. So $\det{g}$ in the harmonic gauge is independent of the mass distribution due to the equations of motion. This means that the space-time geometry around a spherical static mass distribution holds 
\begin{equation}\label{gttgrrr=1}
g_{tt}(r) g_{rr}(r) = -1\,,
\end{equation}  
where $g_{tt}$ and $g_{rr}$ represent respectively the $tt$-component and $rr$-component of the  metric in the standard spherical coordinates.  We, therefore, implicitely consider models of modified gravity wherein  the area-radius coordinate of their spherical-static solution is an affine parameter on the radial null geodesics \cite{Jacobson:2007tj}. 
 
 The simplest non-linear Lagrangian density for $\bar{A}$ preserving \eqref{U1symmetry} and leading to second-order differential equations is
 \begin{eqnarray}\label{315}
{\cal L}_{MOND} & = & -\tilde{\Large \cal L}\left(\frac{\bar{F}_{\mu \nu} \bar{F}^{\mu\nu}}{4 a_0^2}\right) +16\pi G \bar{A}_{\mu} J^\mu \,,
\end{eqnarray}
which after taking the overall factor of 16 in \eqref{overalfactor}  must coincide to \eqref{SMoG} for $\bar{A}_\mu = (4\Phi, \vec{0})$.
Imposing the consistency between \eqref{SMoG} and \eqref{315}, thus, gives:
\begin{equation}
\tilde{\Large \cal L}(8x) =16 a_0^2 {\cal F}(x)\to   \tilde{\Large \cal L}(x)= 16 a_0^2   {\cal F}(\frac{x}{8})\,.
\end{equation}
The consistency between \eqref{315} and the  AQUAL model \eqref{SMoG} demands
\begin{eqnarray}
\label{SA3}
{\cal L}_{MOND} & = & -16 a_0^2{\Large \cal F}\left(\frac{\bar{F}_{\mu \nu} \bar{F}^{\mu\nu}}{32 a_0^2}\right) +16 \pi G \bar{A}_{\mu} J^\mu \,.
\end{eqnarray}
And the equation of motion of $A_\mu$ reads
\begin{equation}\label{moffat5.5}
\nabla_{\nu}({\cal F}' F^{\nu\mu})= 16 \pi G  J^\mu\,,
\end{equation}
where 
\begin{equation}
{\cal F}'= \frac{{\cal F}(x)}{dx}|_{x= -\frac{\bar{F}_{\mu \nu} \bar{F}^{\mu\nu}}{32 a_0^2}}\,,
\end{equation}
Note that this way of extending MOND to GEM is not generally covariant. Next sections provide a generally covariant realization  of  \eqref{SA3}.

\section{Toward the Gauge Vector-Tensor theory }
\label{section6}
The Bekenstein's Tensor-Vector-Scaler theory \cite{Bekenstein:2004ne} is a covariant realization of the AQUAL theory but  does not reproduce \eqref{315}. The observed gravitomagnetism, however,  strongly constraints  the free parameters of the TeVeS theory \cite{Exirifard:2011vb} . We would like to present a covariant generalization of \eqref{315}. To this aim we assume that a  gauge vector field $B_\mu$ and a pseudo Riemannian metric $g_{\mu\nu}$ govern the dynamics of the space-time geometry.  We presume that the orbits of massive particles are derived from the variation of 
\begin{equation}\label{moffat6.1}
S= -m \int d\tau \left(\sqrt{ -\dot{x}^\mu \dot{x}^\mu g_{\mu\nu}(x)}+  B_\mu \dot{x}^\mu \right)\,,
\end{equation}
where $\tau$ is a  parameter defined on the world-line. Eq. \eqref{moffat6.1} is tantamount to saying that the physical length and time are defined in term of a Finsler/Randers geometry \cite{Randers} of
\begin{subequations}
\label{moffat6.1a}
\begin{eqnarray}
S[\tau]&=& -\int d\tau L(x,\dot{x})\,,\\
L(x,\dot{x}) &=& \sqrt{- \dot{x}^\mu \dot{x}^\mu g_{\mu\nu}(x)}+  B_\mu(x) \dot{x}^\mu \,.
\end{eqnarray}
\end{subequations}
The dark matter  and energy   problems  are addressed within the Finsler geometry   \cite{Chang:2008yv, Chang:2009pa, Li:2012qga}. In our setup,  eq. \eqref{moffat6.1} introduces a bi-geometric description for nature where the physical geometry is Finslerian while the geometrical quantities are Riemannian.  

Eq. \eqref{moffat6.1} is the interaction considered in the Moffat's Scalar-Tensor-Vector theory \cite{Moffat:2005si}. We, therefore, adapt  the notation of \cite{Moffat:2005si}.  Let the  Vielbein $e(\tau)$ be introduced on the world-line of the particle \eqref{moffat6.1}: 
\begin{equation}\label{moffat6.b}
S= - \int d\tau \left(-\frac{1}{2} e \dot{x}^\mu \dot{x}^\mu  g_{\mu\nu}+ \frac{m^2}{2e}+ m B_\mu \dot{x}^\mu  \right)\,.
\end{equation}
Parametrizing the world-line such that  $e(\tau)=m$ gives:
\begin{equation}\label{moffat6.c}
S= m \int d\tau \left(\frac{1}{2}  \dot{x}^\mu \dot{x}^\mu  g_{\mu\nu}- B_\mu \dot{x}^\mu  \right)\,,
\end{equation} 
where $\tau$ now is  an affine parameter. Eq. \eqref{moffat6.c} describes the motion  a particle with mass $m$ and an electric charge of $m$ for the $B_\mu$ field.  We will construct the theory such that the contribution of $B_\mu$ to the orbits of particles coincides to that derived from \eqref{315}. Our action takes the form  
\begin{subequations}
\label{GVT}
\begin{equation}
S = S_{\text{Grav}}+ S_{\text{B}}+S_{\text{M}} \,,
\end{equation}
where
\begin{eqnarray}
S_{\text{Grav}} &=& \frac{1}{16 \pi G} \int d^4 x  \, \sqrt{-g} R \,, \\
S_{\text{B}} & =& -\frac{1}{16 \pi G\kappa l^2} \int d^4x \sqrt{- g}\, {\cal L}(\frac{l^2}{4} B_{\mu\nu} B^{\mu\nu})\,,
\end{eqnarray}
\end{subequations}
where $\kappa$ is constant number, $l$ is a constant parameter, $R$ is the Ricci scalar constructed out from $g_{\mu\nu}$  and $B_{\mu\nu}$ is the field strength of $B_\mu$:
\begin{equation}
B_{\mu\nu} = \partial_\mu B_\nu - \partial_\nu B_\mu\,,
\end{equation}
and $S_{\text{M}}$ is the matter's action. The energy momentum tensor is given by 
\begin{equation}
T_{\mu\nu}= T_{M\mu\nu} + T_{B\mu\nu}\,,
\end{equation}
where $T_{M\mu\nu}$ and $T_{B\mu\nu}$ denote respectively the ordinary matter energy-momentum tensor and the energy-momentum tensor contribution of the $B_\mu$ field.  We have
\begin{eqnarray}
T_{M\mu\nu} &\equiv & - \frac{2}{\sqrt{- g}} \frac{\delta S_M}{\delta g^{\mu\nu}} \,,\\
T_{B\mu\nu} &\equiv& - \frac{2}{\sqrt{- g}} \frac{\delta S_B}{\delta g^{\mu\nu}} \,.
\end{eqnarray}
The calculation results:
\begin{equation}
T_{B\mu\nu} = \frac{1}{16 \pi G \kappa} \left({\cal L}' B_{\mu}^{~\alpha} B_{\nu \alpha} - \frac{1}{l^2} g_{\mu\nu}  {\cal L} \right)\,,
\end{equation}
where 
\begin{eqnarray}
 {\cal L} &\equiv & {\cal L}(\frac{l^2}{4 } B_{\mu\nu} B^{\mu\nu}) \,,\\
 {\cal L}' &\equiv & \frac{d  {\cal L}}{ dx}|_{x=\frac{l^2}{4 } B_{\mu\nu} B^{\mu\nu} }\,.
\end{eqnarray}
The matter current density $J^{\mu}$ is defined in terms of the matter action $S_M$:
\begin{equation}
J^{\mu} =  \frac{1}{\sqrt{-g}} \frac{\delta S_M}{\delta B_\mu}\,,
\end{equation}
The metric field equation then follows
\begin{equation}\label{6.13moffat}
G_{\mu\nu} = 8 \pi G\, T_{\mu\nu}\,,
\end{equation}
where $G_{\mu\nu}= R_{\mu\nu}-\frac{1}{2} g_{\mu\nu} R$. The variation of the action with respect to $B_\mu$ gives its equation of motion:
\begin{eqnarray}\label{Moffat6.14}
\nabla_\nu ({\cal L}' B^{\nu\mu}) & =& 16  \pi  G \kappa  J^\mu\,,
\end{eqnarray}
where
\begin{equation}
J^{\mu} = \rho u^\mu\,,
\end{equation}
where $\rho$ is the matter density and $u^\mu$ is its four velocity vector. \eqref{Moffat6.14} is consistent with \eqref{moffat6.1}. It is also similar to \eqref{moffat5.5}.
Our theory resembles the  Moffat's Scalar-Tensor-Vector theory to some extends. However, in contradiction to the Moffats's theory, it is a gauge theory. We also have introduced neither a mass term nor a potential term for the gauge field. Besides no scalars exist. 

Redefining the components of  metric ($g_{ab}$) by \eqref{moffat2.8} and taking the variation of \eqref{moffat6.c} with respect to $x^\mu$ identifies the physical gravitoelectromagnetic fields of our theory:
\begin{equation}
\label{moffatAPhy}
A_\mu^{\text{Phy}} = A_{\mu} + B_{\mu}\,. 
\end{equation}   
Note that $A_\mu^{\text{Phy}}$ is called the physical GEM because it affects the orbits of slow moving massive particles. 

We impose the following asymptotic behaviors on $\cal L$ :
\begin{equation}
\label{617mo}
{\cal L}(x) = \left\{
\begin{array}{ccc}
x & ,&\text{for}~ x \gg 1 \\
\frac{2}{3}|x|^{\frac{3}{2}} &,& \text{for}~ x \le 1
\end{array}
\right.~,
\end{equation}
which is similar to \eqref{moffat413}. 
Let us first look at the solution in the regime of $x\gg 1$ where \eqref{Moffat6.14} simplifies to 
\begin{equation}\label{moffat618}
\nabla_\nu B^{\nu\mu} = 16 \pi G \kappa J^{\mu}
\end{equation}
whose static solutions can be expressed in term of  the GEM approximation to the Einstein-Hilbert gravity, solutions of \eqref{SA2}:
\begin{subequations}
\label{BEHSolar}
\begin{eqnarray}
B_0 &=& 4 \kappa \Phi_{EH} \,,\\
B_i & =&\kappa \vec{A}_{EH} \,.
\end{eqnarray}
\end{subequations}
The extra factor of $4$ in $B_0$ is due to the factor of four in \eqref{moffat4.1}. We assume  that 
\begin{equation}
\label{mo621}
\kappa < 1\,.
\end{equation}
This allows us to neglect the contribution of  the $B$ field to the energy momentum tensor in \eqref{6.13moffat}. This, then, leads to:
\begin{subequations}
\begin{eqnarray}
{A}_0 & = & \Phi_{EH} \,,\\
{A}_i & =& \vec{A}_{EH}\,.
\end{eqnarray}
\end{subequations}
The physical quantities defined in \eqref{moffatAPhy} thus read:
\begin{eqnarray}
\Phi^{\text{Phy}} & = & (1+ 4 \kappa) \Phi_{EH} \,,\\
A_i^{\text{Phy}} & =& (1+\kappa)  \vec{A}_{EH}\,.
\end{eqnarray}
where $\Phi^{\text{Phy}}\equiv A_0+ B_{0}$. Note that $\Phi^{\text{Phy}}$ is read from \eqref{moffat6.c} for $x^{0}=t=\tau$, $x^{i}= cte$.
The Newton's constant is measured by the $\frac{1}{r^2}$ behavior of $\Phi^{\text{Phy}}$. The observed value of the Newton's constant is:
\begin{equation}
G^{\text{obs}} = (1+4 \kappa) G \,.
\end{equation}
Expressing the gravitoelectric and magnetic field in term of the observed value  of the Newton's constant we reach to
\begin{subequations}
\label{moffatphs6.28}
\begin{eqnarray}
\Phi^{\text{Phy}} & = &  \Phi_{EH} \,,\\
A_i^{\text{Phy}} & =& \frac{1+\kappa}{1+4\kappa}  \vec{A}_{EH}\,.
\end{eqnarray}
\end{subequations}
where it is understood that $G^{\text{OBs}}$ replaces $G$. Since ref. \cite{Murphy:2007nt} reports that the measured gravitomagnetic field  is in agreement with the prediction of the Einstein-Hilbert gravity with the precision of $0.1\%$, we  demand that   
\begin{equation}
\label{limitonk}
|\kappa| <  3\times 10^{-4}\,,
\end{equation}
which is consistent with our previous assumption in \eqref{mo621}. 

\section{The  Gauge Vector Tensor Theory}
\label{section7}
The very small lower bound  of $|k|$ in \eqref{limitonk} suggests that we can not consistently describe  nature with only one scale.  In order to have a theory  free of very small constant couplings, we introduce an additional gauge field represented by $\tilde{B}_\mu$:
\begin{subequations}
\label{GVTBB}
\begin{equation}
\label{BBaction}
S = S_{\text{Grav}}+ S_{\text{B}}+S_{\tilde{\text{B}}}+S_{\text{M}} \,,
\end{equation}
where
\begin{eqnarray}
S_{\text{Grav}} &=& \frac{1}{16 \pi G} \int d^4 x  \, \sqrt{-g} R \,, \\
S_{\text{B}} & =& -\frac{1}{16 \pi G\kappa l^2} \int d^4x \sqrt{- g}\, {\cal L}(\frac{l^2}{4} B_{\mu\nu} B^{\mu\nu})\,,\\
S_{\tilde{\text{B}}} & =& -\frac{1}{16 \pi G\tilde{\kappa} \tilde{l}^2} \int d^4x \sqrt{- g}\, \tilde{\cal L}(\frac{\tilde{l}^2}{4} \tilde{B}_{\mu\nu} \tilde{B}^{\mu\nu})\,,
\end{eqnarray}
where $\tilde{B}_{\mu\nu}$ is the field strength of $\tilde{B}_{\mu}$:
\begin{eqnarray}
\tilde{B}_{\mu\nu} = \partial_\mu \tilde{B}_\nu - \partial_\nu \tilde{B}_\mu
\end{eqnarray}
And the orbits of massive particles are derived from
\begin{equation}\label{BB631}
S= m \int d\tau \left(\frac{1}{2}  \dot{x}^\mu \dot{x}^\mu  g_{\mu\nu}- (B_\mu+\tilde{B}_\mu) \dot{x}^\mu  \right)\,.
\end{equation}
\end{subequations} 
Note that $l$ and $\tilde{l}$ are parameters of the theory. We assume that 
\begin{equation}
\label{tllargerl}
\tilde{l} > l\,.
\end{equation}
We also simplify the theory by setting 
\begin{equation}
\label{7.5}
\tilde{\cal L}(x)\equiv  {\cal L}(x)\,,
\end{equation}
while the asymptotic behavior of $\cal L$ is given in \eqref{617mo}. Notice that we  could have chosen 
\begin{equation}
\label{7.6}
\tilde{\cal L}(x)\equiv  x\,,
\end{equation}
However note that \eqref{7.6} can be obtained from \eqref{7.5} by taking the limit of $\frac{l}{\tilde{l}} \to 0$.  Also 
 notice that $\kappa$ and $\tilde{\kappa}$ are  coupling constants of the theory. From this time on, we refer to \eqref{GVTBB} as the GVT theory.

The equations of motion of the Gauge fields follow from the variation of \eqref{BBaction} with respect to $B$ and $\tilde{B}$:
\begin{eqnarray}
\nabla_\nu ({\cal L}' B^{\nu\mu}) & =& 16  \pi  G \kappa  J^\mu\,,\\
\nabla_\nu (\tilde{\cal L}' \tilde{B}^{\nu\mu}) & =& 16  \pi  G \tilde{\kappa}  J^\mu\,,
\end{eqnarray}
where the same matter current is coupled to the gauge fields due to \eqref{BB631}.
Repeating the steps done in the previous section shows that the the GravitoElectroMagnetism approximation to the Newtonian regime of the  GVT theory  receives contribution from  $B$ and $\tilde{B}$ fields:
\begin{subequations}
\label{BB7.8}
\begin{eqnarray}
\Phi^{\text{Phy}} & = & (1+ 4 (\kappa+\tilde{\kappa})) \Phi_{EH} \,,\\
A_i^{\text{Phy}} & =& (1+\kappa+\tilde{\kappa})  \vec{A}_{EH}\,,
\end{eqnarray}
\end{subequations}
where
\begin{equation}
\label{BBPhys7.9}
A_\mu^{\text{Phy}} = A_{\mu}^{EH} + B_{\mu} + \tilde{B}_\mu\,,
\end{equation}   
and the gauge fields solve:
\begin{subequations}
\label{BBstrong}
\begin{eqnarray}
\nabla_\nu B^{\nu\mu} &=& 16 \pi G \kappa J^{\mu}\,\to B_\mu = k \bar{A}^{EH}_\mu\,,\\
\nabla_\nu \tilde{B}^{\nu\mu} &=& 16 \pi G \tilde{\kappa} J^{\mu}\, \to B_\mu = \tilde{k} \bar{A}^{EH}_\mu\,.
\end{eqnarray}
\end{subequations}
We set
\begin{equation}
\label{k=-k}
\kappa + \tilde{\kappa} \equiv 0\,,
\end{equation}
and make the GVT theory consistent with the Einstein-Hilbert prediction.

\section{Regimes of the GVT theory}
\label{section8}
The GVT theory admits the following three regimes:
\subsection{Strong and Newtonian limit}
Eq. \eqref{BBstrong} governs the dynamics of the gauge fields in the strong limit of the GVT theory. We always assume the same boundary conditions on the gauge fields. Eq. \eqref{k=-k} then results   
\begin{equation}
\tilde{B}_\mu = - {B}_\mu\,.
\end{equation}
In other words, we enforce that $B_{\mu}+\tilde{B}_\mu=0$ in the Newtonian and strong regime of the theory. 
We further notice that the contributions of the $B_\mu$ and $\tilde{B}_\mu$ to the energy momentum tensor cancel each other. The strong limit of the theory, therefore, coincides to the Einstein-Hilbert theory. The GVT theory is consistent with all the tests of gravity in the Newtonian and strong regimes.  
 
 \subsection{MOND regime of the GVT theory}
We define the MOND regime of the GVT theory by 
\begin{eqnarray}
{\cal L}(\frac{l^2}{4} B_{\mu\nu} B^{\mu\nu}) &=& \left|\frac{l^2}{4} B_{\mu\nu} B^{\mu\nu}\right|^\frac{3}{2} \,,\\
\tilde{\cal L}(\frac{\tilde{l}^2}{4} \tilde{B}_{\mu\nu} \tilde{B}^{\mu\nu}) &=& \frac{\tilde{l}^2}{4} \tilde{B}_{\mu\nu} \tilde{B}^{\mu\nu} \,.
\end{eqnarray}
 Due to \eqref{tllargerl}, this regime occurs after the Newtonian one. The equations of motion of the gauge fields in the MOND regime simplify to:
\begin{eqnarray}
\label{BB8.4}
\nabla_\nu \left(|B_{\alpha\beta}B^{\alpha\beta}|^{\frac{1}{2}} B^{\nu\mu}\right) &=&\frac{2\kappa}{l} 16 \pi G J^{\mu} \,,\\
\label{BB8.5c}
\nabla_\nu \tilde{B}^{\nu\mu} &=& -16 \pi G \kappa J^{\mu}\,.
\end{eqnarray}
In this regime the physical gravitoelectric field reads 
\begin{equation}
\Phi^{\text{Phy}}= \Phi+ B_{0} + \tilde{B}_0 \approx B_0\,.
\end{equation} 
where $\Phi$ is produced by the $tt$ component of the metric.  
In the absence of the gravitomagnetic field ($B_i=0$), the eq. \eqref{BB8.4} converts to
\begin{equation}
\label{6.34mo}
\nabla^i \left(|\nabla \Phi^{\text{Phy}}| \nabla_i \Phi^{\text{Phy}}\right) =\frac{ 4\sqrt{2} \kappa}{l} 4 \pi G \rho \,.
\end{equation}
where $B_{\alpha\beta}B^{\alpha\beta} = 2 |\nabla \Phi^{\text{Phy}}|^2$ is used.
The consistency between \eqref{phiMOND0} and \eqref{6.34mo} demands that 
\begin{equation}\label{a0k9.7}
a_0 = \frac{4\sqrt{2}\kappa c^2}{l}\,,
\end{equation}
wherein the dependency on $c$ is recovered. Notice that the MOND regime starts when 
\begin{equation}
\frac{l^2}{4} B_{\mu\nu} B^{\mu\nu} \le 1\,.
\end{equation} 
This is where the Newtonian regime ends. In the Newtonian regime of a stationary mass distribution $B_\mu=\kappa \bar{A}_\mu=\kappa(4 \Phi_{N},0)$ where $\Phi_N$ presents the Newtonian potential. Therefore the  Newtonian regime ends at
\begin{equation}
|\nabla \Phi_{N}| < \frac{a_0}{16 \kappa^2}\,.
\end{equation}
where \eqref{a0k9.7} is used to express $l$ in term of $a_0$ and the dependency on $c$ is recovered. This means that the MOND regime occurs in 
\begin{equation}
4 |\kappa| \sqrt{\frac{GM}{a_0}} < r \,.
\end{equation}
We assume that $\kappa= O(1)$ in order to keep the GVT theory consistent with observations. In particular we note that for
\begin{equation}
\kappa= \pm \frac{1}{4}\,,
\end{equation}
the boundary of the MOND regime of the GVT theory coincides to that of the AQUAL theory. Let it be highlighted that \eqref{limitonk} contradicts observations in the Solar system. In order to avoid such a contradiction, we have introduced two gauge fields rather than only one.  
 \subsection{Post-MONDian limit}
 We define the Post-MONDian regime of the GVT theory by 
\begin{eqnarray}
{\cal L}(\frac{l^2}{4} B_{\mu\nu} B^{\mu\nu}) &=& \left|\frac{l^2}{4} B_{\mu\nu} B^{\mu\nu}\right|^\frac{3}{2} \,,\\
\tilde{\cal L}(\frac{\tilde{l}^2}{4} \tilde{B}_{\mu\nu} \tilde{B}^{\mu\nu}) &=& \left|\frac{\tilde{l}^2}{4} \tilde{B}_{\mu\nu} \tilde{B}^{\mu\nu}\right|^\frac{3}{2} \,.
\end{eqnarray}
 Due to \eqref{tllargerl}, this regime occurs after the MOND regime when
\begin{equation}
\label{PN8.14}
\frac{\tilde{l}^2}{4} \tilde{B}_{\mu\nu} \tilde{B}^{\mu\nu} \le 1\,.
\end{equation}
The equations of motion of the gauge fields in the Post-MONDian regime simplify to:
\begin{eqnarray}
\nabla_\nu \left(|B_{\alpha\beta}B^{\alpha\beta}|^{\frac{1}{2}} B^{\nu\mu}\right) &=&\frac{2\kappa}{l} 16 \pi G J^{\mu} \,,\\
\nabla_\nu \left(|\tilde{B}_{\alpha\beta}\tilde{B}^{\alpha\beta}|^{\frac{1}{2}} \tilde{B}^{\nu\mu}\right)  &=& -\frac{2\kappa}{\tilde{l}} 16 \pi G J^{\mu}\,,
\end{eqnarray}  
We see that  the $\tilde{B}_\mu$ field contributes to the Post-MONDian regime. The behavior of the $\tilde{B}_\mu$ is like that of ${B}_\mu$ but rescaled and with a negative sign. Let it be defined:
\begin{eqnarray}
\tilde{a}_0 &\equiv& \beta^2 a_0\,,\\
\label{820beta}
 \beta &\equiv & \sqrt{\frac{l}{\tilde{l}}}\,.
\end{eqnarray}
Before the start of the post-MONDian regime the $\tilde{B}_\mu$ fields solves \eqref{BB8.5c}. So around a spherical stationary solution $B_0 = -4\kappa\Phi_N$. The condition of \eqref{PN8.14} then implies that  the  post-MONDian regime starts at
\begin{equation}
\label{PN8.20}
|\nabla \Phi_{N}| < \frac{\tilde{a}_0}{16 \kappa^2}\,,
\end{equation}
and continues to infinity. 

\section{Gravitomagnetism  of a  spherical mass distribution in the GVT theory}
\label{section9} 
This section studies the gravitomagnetism produced by a spherical static mass distribution in the three regimes of the GVT theory.
\subsection{Newtonian regime}
The gravitoelectric and gravitomagnetic fields that a slow rotating spherical mass distribution produce  in the Newtonian regime follow from  \eqref{BB7.8} and \eqref{k=-k}:
\begin{subequations}
\label{6162B2}
\begin{eqnarray}
-\nabla \Phi^{Phy} & = & -\frac{G M}{r^3}\vec{r} \,,\\
\nabla \times A_{Phys} & = &\nabla \times A_{EH}=\frac{G}{2 c^2} (\frac{J}{r^3} - 3 J.r \frac{\vec{r}}{r^5})\,,
\end{eqnarray}
\end{subequations}
where $M$ is the total mass and $J$ is the total angular velocity of the spherical mass,  $\vec{r}= 0$ represents the center of the mass distribution and $G$ is the Newton's constant.
\subsection{MONDian regime}
Identifying   $\bar{A}_\mu$ , $\bar{B}_\mu$ and $B_{\mu}$ fields inside the MOND windows precedes  the physical GEM. In this regime, the equations for $A_\mu$ and  $\bar{B}_\mu$ are those of the Einstein-Hilbert theory. So:
\begin{subequations}
\label{BB9.2}
\begin{eqnarray}
-\nabla A_0 & = & -\frac{G M}{r^3}\vec{r} \,,\\
\nabla \times A_{i} & = &=\frac{G}{2 c^2} (\frac{J}{r^3} - 3 J.r \frac{\vec{r}}{r^5})\,,
\end{eqnarray}
\end{subequations}
and 
\begin{subequations}
\label{BB9.3}
\begin{eqnarray}
-\nabla \tilde{B}_0 & = & 4 \kappa \frac{G M}{r^3}\vec{r} \,,\\
\nabla \times \tilde{B}_{i} & = &=-\kappa \frac{G}{2 c^2} (\frac{J}{r^3} - 3 J.r \frac{\vec{r}}{r^5})\,.
\end{eqnarray}
\end{subequations}
The eq. \eqref{BB8.4}, being the equation of motion of $B_\mu$ field in the MOND regime, simplifies to
\begin{subequations}
\label{9.4BBc}
\begin{eqnarray}
\nabla^i \left(\frac{\sqrt{|\nabla B_0|^2- |\nabla \times \vec{B}|^2}}{a_0}\nabla_i B_0\right) &=& 4 \pi G \rho\,,\\
-\nabla \times \left(\frac{\sqrt{|\nabla B_0|^2- |\nabla \times \vec{B}|^2}}{a_0} \nabla \times \vec{B}\right) &=& 4\pi G \vec{J}\,,
\end{eqnarray}
\end{subequations}
where $B_\mu=(B_0,\vec{B})$. Because a slow rotating mass distribution holds $ |\nabla A_0|^2\ll |\nabla \times \vec{A}|^2$, \eqref{9.4BBc} can be approximated to:
\begin{subequations}
\begin{eqnarray}
\nabla^i \left(\frac{|\nabla B_0|}{a_0}\nabla_i B_0\right) &=& 4 \pi G \rho\,,\\
-\nabla \times \left(\frac{|\nabla B_0|}{a_0} \nabla \times \vec{B}\right) &=& 4\pi G \vec{J}\,,
\end{eqnarray}
\end{subequations}
whose solutions can be expressed in terms of the Einstein-Hilbert GEM:
\begin{eqnarray}\label{EMondmof}
\frac{|\nabla B_0|}{a_0} \nabla B_0  & = & \nabla\Phi_{EH} + \nabla\times \vec{\tilde{h}}\,, \\
\label{BMondmof}
\frac{|\nabla B_0|}{a_0}\nabla\times  \vec{B} & = &\frac{1}{4} \left(\nabla\times \vec{A}_{EH} + \nabla \tilde{h}\right)\,,
\end{eqnarray}
where $\vec{\tilde{h}}$ and $\tilde{h}$ solve
\begin{eqnarray}\label{eqhemof}
0\,=\,\nabla \times \nabla B_0  & = & \nabla\times(\frac{ \nabla\Phi_{EH} + \nabla\times \vec{\tilde{h}}}{|\nabla B_0|})\,,\\\label{eqhbmof}
0\,=\, \nabla.\nabla\times  \vec{B} & = &\nabla.( \frac{\nabla\times A_{EH} + \nabla \tilde{h}}{|\nabla B_0|})\,.
\end{eqnarray}
Since  $\vec{\tilde{h}}=0$    solves \eqref{eqhemof} then 
\begin{equation}\label{8.9mof}
\nabla B_0 = \frac{\sqrt{GM a_0}}{r^2}\vec{r}\,.
\end{equation}
Inserting  \eqref{8.9mof} into the consistency equation for $\nabla{\tilde{h}}$ yields
\begin{equation}\label{conditiononh}
\nabla. (r \nabla \tilde{h})  \,=\, \frac{G}{4c^2 (G M a_0)^{\frac{1}{2}} } \frac{J.r}{r^4}\,,
\end{equation}
which is a non-homogeneous Laplace's equation in four dimensions written in the spherical coordinates:
\begin{equation}\label{Laplace4}
\Box_{4} \tilde{h} = \frac{G}{c^2 (G M a_0)^{\frac{1}{2}} }  \frac{J. r}{r^5}\,,
\end{equation}
 where $\tilde{h}\equiv\tilde{h}(r, r.J)$ is understood.  Let it be emphasized that \eqref{Laplace4} represents the $\tilde{h}$ equation in large $r$. It holds $\nabla\tilde{h}=\vec{0}$ near the origin.  We choose a solution of \eqref{Laplace4}  which is source free  at the origin. Doing so, the fall off of the $\nabla \tilde{h}$ is guaranteed to be $r^{-4}$ or less.  $\nabla\times B$ in the MOND regime, therefore, yields 
\begin{equation}\label{AmondSp}
\nabla \times B \,= \,  - \frac{G}{8c^2(G M  a_0)^\frac{1}{2}}  (\frac{J}{r^2} - 3 J.r \frac{\vec{r}}{r^4}) + O(\frac{1}{r^4})\,.
\end{equation}
Note that \eqref{AmondSp} is not divergent for  small masses  because $|J| \propto M$. 
The physical GEM in the MOND regime follows from \eqref{BBPhys7.9}, \eqref{BB9.3} and \eqref{BB9.2}:
\begin{eqnarray}
\label{713mo}
-\nabla \Phi^{\text{Phy}}&=& -\left((1-4 \kappa)\frac{G M}{r^3} +\frac{\sqrt{GM a_0}}{r^2}\right)\vec{r}\,,\\
\label{714mo}
\nabla \times \vec{A}^{\text{Phy}} &=&  \frac{G}{8c^2(G M  a_0)^\frac{1}{2}}  (\frac{J}{r^2} - 3 J.r \frac{\vec{r}}{r^4}) +\,(1-\kappa)  \, \frac{G}{2 c^2} (\frac{J}{r^3} - 3 J.r \frac{\vec{r}}{r^5})+ O(\frac{1}{r^4})\,.
\end{eqnarray}
\begin{figure}[tbp]
        \centering
        \begin{subfigure}[b]{0.45\textwidth}
                \centering
                \includegraphics[width=\textwidth]{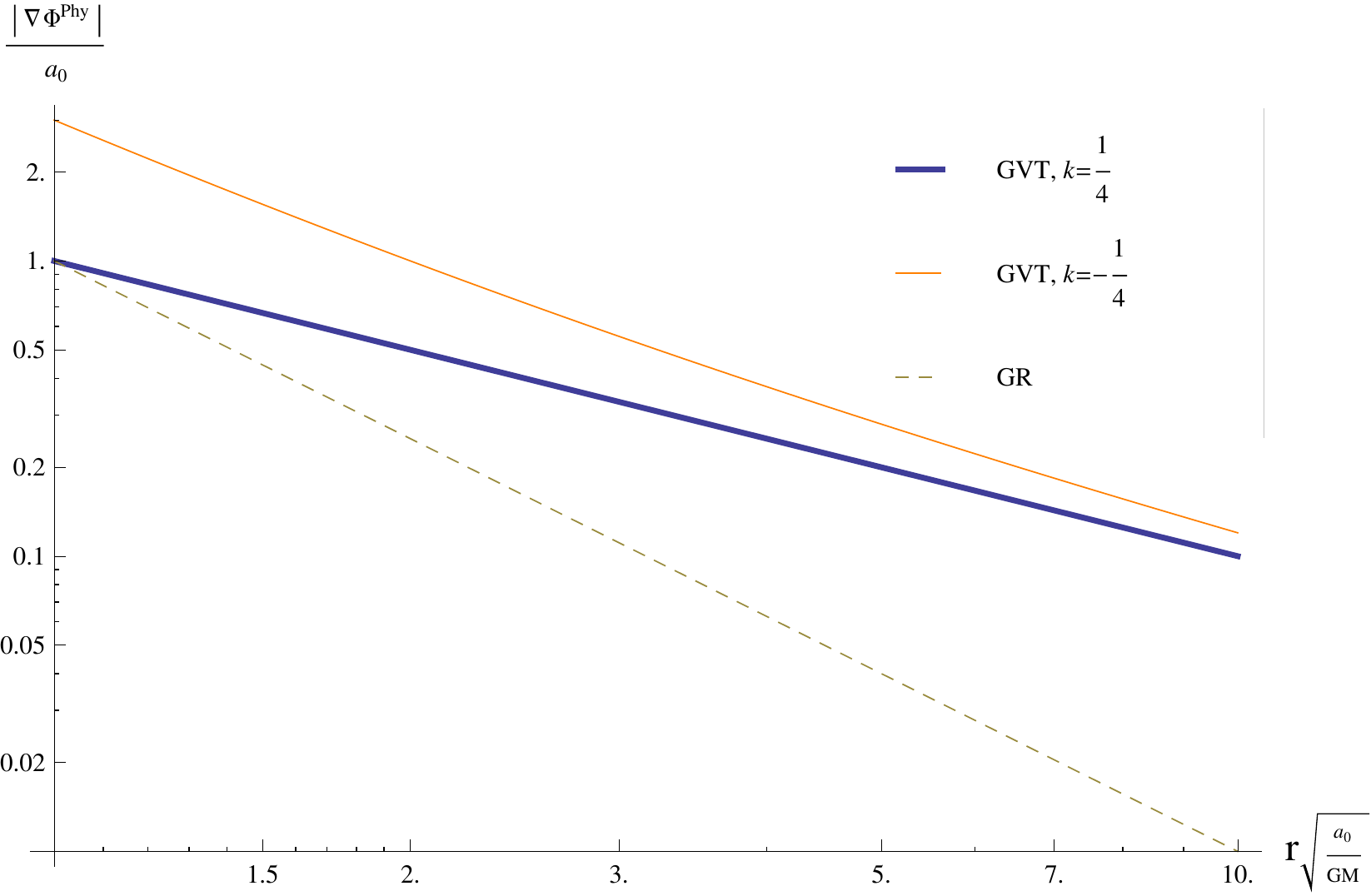}
                \caption{The GravitoElectric Field strength}
                \label{fig:gull}
        \end{subfigure}%
        ~ 
        \begin{subfigure}[b]{0.45\textwidth}
                \centering
                \includegraphics[width=\textwidth]{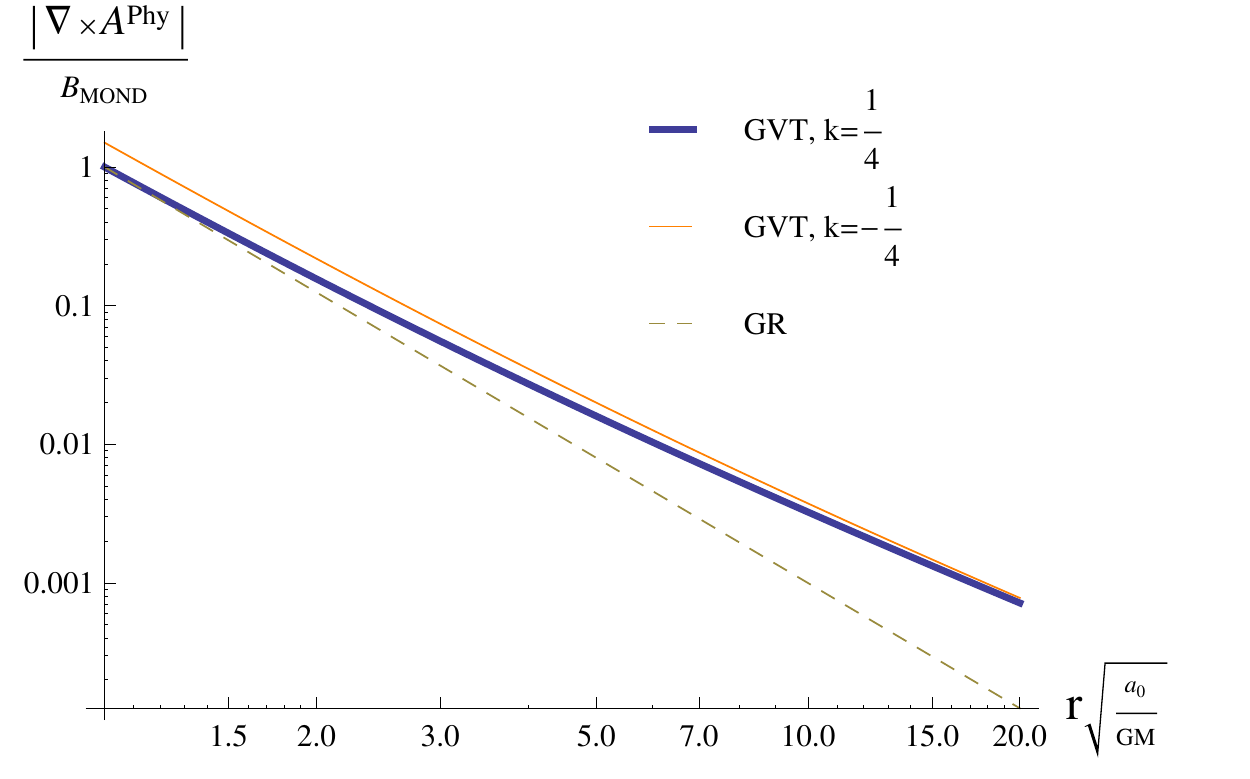}
                \caption{The GravitoMagnetic Field strength}
                \label{fig:tiger}
        \end{subfigure}
        \caption{The GravitoElectroMagnetism of a spherical static mass distribution in the Einstein-Hilbert theory (GR)  and the Gauge Vector Tensor theory (GVT).}
        \label{fig:1}
\end{figure}%
Eq. \eqref{713mo} for $\kappa=\frac{1}{4}$  is the ordinary MOND modification of the Newtonian field capable of resolving the missing mass problem in galaxies and reproducing the Tully-Fisher relation \cite{Tully}. This suggests to set 
\begin{equation}
\kappa = \frac{1}{4}\,.
\end{equation}
Fig. \ref{fig:1} depicts the magnitude of \eqref{713mo}, and the magnitude of \eqref{714mo} for $J.r=0$ for two values of $k$ and
\begin{equation}
B_{MOND}= \frac{G |J| a_0^{\frac{3}{2}}}{2 c^2 ( G M)^{\frac{3}{2}}}\,.
\end{equation}
We see that the fall off of the  gravitomagnetic field strengths of GVT in its MOND regime is $r^{-2}$ while that of the Einstein-Hilbert theory is $r^{-3}$. The gravitomagnetic field is enhanced in the deep MOND regime.

The equations for the Newtonian potential and the gravitomagnetic field  of  the $\Lambda$CDM theory read
\begin{eqnarray}\label{phiCDM}
\nabla^2 \Phi & =& 4\pi G (\rho+\rho_{\text{Dark}} ) \,,\\ \label{gravoCDM}
\nabla^2 \vec{A} & =& \frac{16 \pi G}{c^2} (\vec  j+ \vec{j}_{\text{Dark}} )\,,
\end{eqnarray}
where $\rho_{\text{Dark}}$ and $j_{\text{Dark}}$ are respectively the density and the angular velocity distributions of dark matter.  The gravitomagnetic field strength that the $\Lambda$CDM theory predicts for a spherical spinning galaxy at its edge then follows
\begin{eqnarray}\label{CDMgravo}
\nabla\times \vec{A}_{\Lambda\text{CDM}} & =&  \frac{G}{2 c^2} (\frac{J}{r^3} - 3 J.r \frac{\vec{r}}{r^5})\,.
\end{eqnarray}
There exists no observational information available about the  angular momentum distribution of dark matter. The theoretical scenarios consider the  dark matter halo as a cloud of a vanishing angular momentum \cite{Bhattacharjee:2012xm}.  We, additionally, observe that  the difference between GVT \eqref{714mo}  and $\Lambda$CDM \eqref{CDMgravo}  can not be  assigned to the total angular momentum of the dark matter.  We, therefore, conclude that  measuring the gravitomagnetic force at the edge/beyond the edge of a  galaxy  refutes one of the GVT and dark paradigms and proves the other one. However the gravitomagnetic force at the edge of a galaxy is too small  that one may not hope for its detection in the near future. 

\subsection{Post-MONDian regime}
Due to \eqref{PN8.20} the post-MONDian field starts from
\begin{equation}
\frac{4\kappa}{\beta} \sqrt{\frac{GM}{a_0}} < r\,. 
\end{equation}
and continues to infinity. In the post-MONDian regime, the $\tilde{B}_\mu$ field starts to behave like the $B_\mu$ field. So the physical gravitomagnetism in this regime follows:
\begin{eqnarray}
-\nabla \Phi^{\text{Phy}}&=& -\left(\frac{G M}{r^3} +(1-\beta)\frac{\sqrt{GM a_0}}{r^2}\right)\vec{r}\,,\\
\nabla \times \vec{A}^{\text{Phy}} &=&  (1-\beta ) \frac{G}{8c^2(G M  a_0)^\frac{1}{2}}  (\frac{J}{r^2} - 3 J.r \frac{\vec{r}}{r^4}) + \, \frac{G}{2 c^2} (\frac{J}{r^3} - 3 J.r \frac{\vec{r}}{r^5})+ O(\frac{1}{r^4})\,,
\end{eqnarray}
where $\beta$ is defined in \eqref{820beta}. Since $\beta$ is smaller than one, the MONDian behavior of the GVT though is decreased continues to infinity. 

We note that  the post-MONian behavior of the GVT theory can be enforced  to coincide to  the Newtonian one by introducing one additional gauge field.   Let $\hat{B}_\mu$ be introduced whose action is similar to that of $B_\mu$ where $(\kappa,l)$ is replaced by $ (\hat{\kappa},\hat{l})$. The interested reader can check that 
\begin{equation}
\kappa+\tilde{\kappa}+\hat{\kappa}=0\,,
\end{equation}
makes the theory consistent with the Einstein-Hilbert action in the strong and Newtonian regimes while the condition of   
\begin{equation}
\frac{\kappa}{l}+\frac{\tilde{\kappa}}{\tilde{l}}+\frac{\hat{\kappa}}{\hat{l}}=0\,,
\end{equation}
 causes the theory to be consistent with the Einstein-Hilbert theory in the post-MONDian regime. Such a simple extension indicates to an advantage of the GVT theory over its rivals.
\section{Gravitomagnetic field in the Solar system}
\label{fasle4-2}
In the TeVeS and the AQUAL theories, in some points  within the solar system the gravitational fields of the planets and the Sun and the galaxy cancel each other. Let these points be called the gravitational saddle points. Ref. \cite{Galianni:2011ch} identifies the gravitational saddle points of the solar system. Ref.  \cite{Bekenstein:2006fi,Mozaffari:2011ux,Magueijo:2011an}  suggest that an accurate tracking of a probe like the LISA path finder that passes through the MOND windows can prove or refute the AQUAL theory. Ref. \cite{Exirifard:2012hj} proposes that measuring the behavior of gravity in short distances within the MOND windows can prove or refute the AQUAL theory.  Ref. \cite{Exirifard:2011zk} mentions that observing pulsars through the gravitational saddle point of the Sun and Jupiter can empirically constrain the interaction of light with the physics of MOND system. This section aims to study the physics within the GVT MOND windows of the Solar system. To this aim we  will consider the largest MOND window. The subsection \ref{MONDAQUAL} reviews the MOND window of the AQUAL theory. Then the subsection \ref{GVTMOND} identifies the Sun-Jupiter MOND window of the GVT theory. The subsection \ref{subsection10.3} solves the GVT equations in the Sun-Jupiter MOND window.  

\subsection{MOND windows of the AQUAL theory}
\label{MONDAQUAL}
This section aims to study the MOND windows in the framework of the AQUAL theory.  To this aim we  shall consider the largest solar MOND window. We will consider the gravitational saddle point of the Sun-Jupiter system. 
 We employ the two bodies approximation to the Sun-Jupiter system. This approximation suffices for our studies because including the effects of other solar planets and the gravitational field of the galaxy will not significantly change the size of the considered MOND window \cite{Galianni:2011ch}.   
\begin{figure}[tbp]
        \centering
                \centering
                \includegraphics[width=\textwidth]{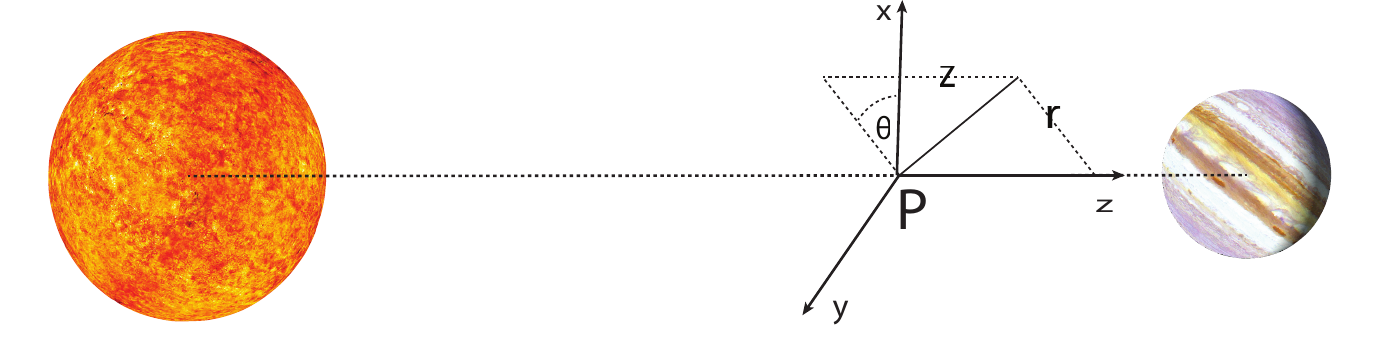}
                \caption{$P$ represents the gravitational saddle point of the Sun-jupiter system. We use the cylindrical coordinates around $P$ in order to solve the equations.}
                \label{fig:2}
\end{figure}
In this approximation the Newtonian gravitational field strength at the position $\vec{r}$ with respect to the center of the Sun reads:
\begin{equation}
\label{BB10.1}
g(\vec{r}) = -\frac{G M_{\text{Sun}}}{|\vec{r}|^3} \vec{r} - \frac{G M_\text{Jupiter}}{|\vec{r}-\vec{d}|^3}(\vec{r}-\vec{d})\,.
\end{equation}
where  $\vec{d}$ is the vector connecting the center of the  Sun to the center of the Jupiter. The gravitational saddle point $P$ is the point where in $g(\vec{r}_p) =0$, so:
\begin{equation}
\label{rp10.2}
\vec{r}_{p} =  \frac{\vec{d}}{1+ \sqrt{\frac{M_{\text{Jupiter}}}{M_{\text{Sun}}}}}\,.
\end{equation}
which means that the saddle point is $2.29 \times 10^{7} km$ far away from the Jupiter. 
We would like to study the physics around the gravitational saddle point. We, therefore, taylor-expand the gravitational field around the saddle point:
\begin{eqnarray}
\vec{r}&=& \vec{r}_{p} + \epsilon \delta r\,, \\
 \delta r &=& \rho \hat{\rho} + z \hat{z}\,,\\
 \label{graqual}
\vec{g}(\vec{r}) &=& 0 +\alpha ( 2 z \hat{z} - \rho \hat{\rho}) +O(z^2, \rho^2)\,, 
\end{eqnarray}
where fig. \ref{fig:2} depicts the chosen cylindrical coordinate and 
\begin{equation}
\alpha\,=\, \frac{G}{d^3}\frac{\left(\sqrt{M_{\text{Sun}}} +  \sqrt{M_{\text{Jupiter}}}\right)^4}{\sqrt{M_{\text{Sun}} M_{\text{Jupiter}}}}=1.084 \times 10^{-14} s^{-2}.
\end{equation}
 The magnitude of the gravitoelectric field strength around the gravitational saddle point then follows:
\begin{equation}
|\vec{g}(\vec{r})| \,=\, 0 +\alpha \epsilon \sqrt{4 z^2  + \rho^2} +O(z^2,\rho^2)\,,
\end{equation}
The AQUAL type MOND window is where 
\begin{equation}
\label{10.8AqualBB}
|\vec{g}(\vec{r})| < a_0.
\end{equation}
So it is an ellipsoid with semi-axes of length 
\begin{subequations}
\label{10.9AQUALBB}
\begin{eqnarray}
a &= &\frac{a_0}{2\alpha}\,=\,4.6~ km \,,\\
b &= &\frac{a_0}{\alpha}\,=\, 9.2~ km\,,\\
c &=&\frac{a_0}{\alpha}\,=\,9.2~ km\,.
\end{eqnarray}
\end{subequations}
\subsection{MOND windows of the GVT theory}
\label{GVTMOND}
The Newtonian regime of the GVT theory ends at
\begin{eqnarray}
\label{10.11}
|\frac{l^2}{4} B_{\mu\nu} B^{\mu\nu}|=\frac{l^2}{2} \left(|\nabla B_0|^2 - |\nabla \times \vec{B}|^2)\right) &=& 1 \,,
\end{eqnarray}
where $B$ is given in \eqref{BEHSolar} and the unit of $c=1$ is used. Utilizing \eqref{BB10.1}, $B_0$ in the Solar system reads 
\begin{equation}
\label{10.12}
\nabla B_0(\vec{r}) =4 k \vec{g}(\vec{r})=-4\kappa \left(\frac{ G M_{\text{Sun}}}{|\vec{r}|^3} \vec{r} + \frac{ G  M_\text{Jupiter}}{|\vec{r}-\vec{d}|^3}(\vec{r}-\vec{d})\right)\,.
\end{equation}
The rotation of the Sun and Jupiter around their axes  as well as the motion of the center of the mass of the Jupiter around the Sun contribute to  the gravitomagnetic field strength in the Solar system:  
\begin{subequations}
\label{GVTB10.13}
\begin{eqnarray}
 \nabla \times \vec{A}_\text{Jupiter}&=& \frac{G}{2 c^2}\left(\frac{\vec{J}_J}{(d-r)^3}- 3 \vec{J}_J.(\vec{d}-\vec{r}) \frac{\vec{d}-\vec{r}}{|\vec{d}-\vec{r}|^5}\right)\,,\\
 \nabla \times \vec{A}_\odot&=& \frac{G}{2 c^2}\left(\frac{\vec{J}_\odot}{r^3}- 3 \vec{J}_\odot.\vec{r} \frac{\vec{r}}{r^5}\right)\,,\\
 \nabla \times \vec{A}_\text{Jupiter}^{\text{Orbital}}&=& \frac{G M_\text{Jupiter}}{ c^2 |\vec{r}-\vec{d}|^3} \vec{v}_{cm}\times (\vec{r}-\vec{d})\,,
\end{eqnarray}
\end{subequations}
where $J_\odot$ and $J_J$ are the angular momentum of the Sun and Jupiter while $v_{cm}$ is the velocity of the center of the mass of Jupiter with respect to the Sun. Utilizing \eqref{BB10.1} now leads to:
\begin{eqnarray}
\label{10.16}
\nabla \times \vec{B}& =& \kappa( 
 \nabla \times \vec{A}_\text{Jupiter}+  \nabla \times \vec{A}_\odot +  \nabla \times \vec{A}_\text{Jupiter}^{\text{Orbital}})\,,
 \end{eqnarray}
 Eq. \eqref{10.11}, \eqref{10.12} and \eqref{10.16} identifies the boundary of the GVT MOND window of the Sun-Jupiter system   in the two bodies approximation to the Solar system. This is the boundary of the MOND window with the Newtonian regime.  Let it be highlighted that including the effects of other solar planets  will not significantly change the size of this MOND window \cite{Galianni:2011ch}.

Since $k\approx O(1)$, the eq. \eqref{a0k9.7} implies that $l \approx O(10^{26} m)$. Recalling that the gravitomagnetic field strength is weaker than the gravitoelectric field strength, \eqref{10.11} then implies that the GVT MOND window is not  far away from the gravitational saddle point given in \eqref{rp10.2}.  To go further we approximate  the angular momentum of the Sun and Jupiter to:
\begin{eqnarray}
J_\odot &=& 1.92\times 10^{41} kg m^2 s^{-1} \hat{x}\,,\\
J_{J} & = & 6.9\times 10^{38}  kg m^2 s^{-1} \hat{x}\,,\\
v_{cm} &=& d \frac{2\pi}{T}\hat{y}\,,
\end{eqnarray}
where $\hat{x}$ and $\hat{y}$ are presented in fig. \ref{fig:2} and $T$ is the orbital period of the Jupiter around the Sun:
\begin{equation}
T= 4331.6~\mbox{days}\,.
\end{equation}
We see that the magnitudes of the gravitomagnetic fields presented in \eqref{GVTB10.13} at $\vec{r}_p$ (and as well as its neighborhood)  given  in \eqref{rp10.2} read:
\begin{subequations}
\label{10.18BB3}
\begin{eqnarray}
c \nabla \times A^{Sun}_{rotation} & = & \frac{G}{2 c} \frac{J_{Sun}}{r^3_p} \hat{x}\,= 1.05  \times 10^{-13} \frac{m}{s^2} \hat{x},\\
c \nabla \times A^{Jupiter}_{rotation} & = & \frac{G}{2 c} \frac{J_{Jupiter}}{(d-r_p)^3} \,=\,1.27 \times 10^{-11} \frac{m}{s^2}\hat{x}\,,\\
c \nabla \times A^{Jupiter}_{\text{orbital}}  &= & \frac{G}{c}\frac{ M_{\text{Jupiter} }} {d-r_p} \frac{2 \pi}{T}\,\hat{z}=3.10 \times 10^{-10}\frac{m}{s^2}  \hat{x}\,,
\end{eqnarray}
\end{subequations}
which in turn result
\begin{equation}
\label{10.22}
c^2 |\nabla \times B|^2 \approx 10.41\,\kappa^2 \,a_0^2\,, 
\end{equation}
where \eqref{valuea0mo} and \eqref{10.16} are used. Inserting \eqref{10.22} into \eqref{10.11} and expressing $l$ in term of $a_0$ by \eqref{a0k9.7}, and utilizing \eqref{10.12} yields
 \begin{equation}
 \label{10.22a}
|g(r)| = a_0 \sqrt{\frac{1}{256 \kappa^4} + 0.65}\,.
 \end{equation}
Notice that  eq. \eqref{graqual} decribes the Newtonian gravitational field strength around $\vec{r}_p$. Now comparing \eqref{10.22} with \eqref{10.8AqualBB} and \eqref{10.9AQUALBB} results that the GVT MOND window is an ellipsoid with semi-major axes of 
 \begin{equation}
 (\tilde{a},\tilde{b},\tilde{c}) =  \sqrt{\frac{1}{256 \kappa^4} + 0.65}~\frac{a_0}{ \alpha} (\frac{1}{2}, 1 ,1)\,.
 \end{equation}
 The GVT MOND window is larger than the AQUAL MOND window for $|k|<0.325$. 
 
 The post MONDian regime resides inside the GVT MOND windows. The post MONDian regime is an ellipsoid with semi axes of 
\begin{equation}
 (\tilde{\tilde{a}},\tilde{\tilde{b}},\tilde{\tilde{c}}) = \,\beta^2 (\tilde{a},\tilde{b},\tilde{c}),
 \end{equation}
 where $\beta$ is given in \eqref{820beta}. Note that $\tilde{l}$ is the scale wherein $\tilde{B}$ starts its MONDian behavior. We assume that $\beta\ll 1$. This makes the post MONDian region of the Solar system  sufficiently small to  practically be ignored.

\subsection{Gravitomagnetism inside the Sun-Jupiter GVT MOND window} 
\label{subsection10.3}
The physical GravitoElectroMagnetism in the GVT theory receives contribution from the metric and the gauge fields, as stated in the eq. \eqref{BBPhys7.9}. The contribution of the metric to GEM inside the GVT MOND windows follows from \eqref{graqual} and \eqref{10.18BB3}:
\begin{subequations}
\label{10.23BBc}
\begin{eqnarray}
-\nabla \Phi_{EH} &=& 0 +\alpha ( 2 z \hat{z} - \rho \hat{\rho}) \,, \\
\nabla \times A_{EH} &=& 3.22 \times 10^{-10}\frac{m}{s^2}  \hat{x}\,.
\end{eqnarray}
\end{subequations}
The contribution of the $\tilde{B}$ follows from \eqref{BB8.5c}:
\begin{subequations}
\label{1024BBc}
\begin{eqnarray}
-\nabla \tilde{B}_0 &=& -4\kappa \alpha  ( 2 z \hat{z} - \rho \hat{\rho})\,, \\
\nabla \times \tilde{B}_i &=& -3.22\kappa \times 10^{-10}\frac{m}{s^2}  \hat{x}\,.
\end{eqnarray}
\end{subequations}
We should solve \eqref{9.4BBc} in order to  find the contribution of $B_\mu$. The solution of \eqref{9.4BBc} can be expressed in term of  \eqref{10.23BBc}:
\begin{subequations}
\label{1026BBc}
\begin{eqnarray}
\frac{\sqrt{| |\nabla B_0|^2- |\nabla \times \vec{B}|^2|}}{a_0} \nabla B_0  & = & -\alpha ( 2 z \hat{z} - \rho \hat{\rho}) + \nabla\times \vec{\tilde{h}}\,, \\
\frac{\sqrt{||\nabla B_0|^2- |\nabla \times \vec{B}|^2|}}{a_0} \nabla\times  \vec{B} & = & 3.22 a_0  \hat{x} + \nabla \tilde{h}\,,
\end{eqnarray}
where $ \vec{\tilde{h}}$ and $\nabla \tilde{h}$ solve the following consistency equations:
\begin{eqnarray}
\nabla \times \nabla B_0 &=&\nabla \times \frac{-\alpha ( 2 z \hat{z} - \rho \hat{\rho}) + \nabla\times \vec{\tilde{h}}}{\sqrt{| |\nabla B_0|^2- |\nabla \times \vec{B}|^2|}}= 0\,,\\
\nabla.\nabla \times \vec{B} &=&\nabla.\frac{3.22 a_0  \hat{x} + \nabla \tilde{h}}{\sqrt{| |\nabla B_0|^2- |\nabla \times \vec{B}|^2|}}=0\,. 
\end{eqnarray}
\end{subequations}
We first look at part of the GVT MOND window wherein 
\begin{equation}
\label{1026BBcGheyd}
 |\nabla \times \vec{B}|^2> |\nabla B_0|^2\,,
\end{equation}
where \eqref{1026BBc} can be approximated to:
\begin{subequations}
\label{10.28BBc}
\begin{eqnarray}
\frac{ |\nabla \times \vec{B}|}{a_0} \nabla B_0  & = & -\alpha ( 2 z \hat{z} - \rho \hat{\rho}) + \nabla\times \vec{\tilde{h}}\,, \\
\frac{ |\nabla \times \vec{B}|}{a_0} \nabla\times  \vec{B} & = & 3.22 a_0  \hat{x} + \nabla \tilde{h}\,,\\
\nabla \times \frac{-\alpha ( 2 z \hat{z} - \rho \hat{\rho}) + \nabla\times \vec{\tilde{h}}}{|\nabla \times \vec{B}|}&=&0\,,\\
\nabla.\frac{3.22 a_0  \hat{x} + \nabla \tilde{h}}{ |\nabla \times \vec{B}|}&=&0\,. 
\end{eqnarray}
\end{subequations}
Eq. \eqref{10.28BBc} is solved by
\begin{subequations}
\label{1028BBc}
\begin{eqnarray}
\nabla \times \vec{B} &=& 1.79 a_0 \hat{x}\,,\\
\nabla {B}_0 &=&- 0.55 \alpha ( 2 z \hat{z} - \rho \hat{\rho}) \,,\\
\label{1028BBcc}
 \tilde{h} &=& \vec{\tilde{h}} =0 \,.
\end{eqnarray}
\end{subequations}
The condition of \eqref{1026BBcGheyd} applied on \eqref{1028BBc} gives 
\begin{equation}\label{10.29BBC}
\sqrt{4z^2+\rho^2} < 3.25 \frac{a_0}{\alpha}\,,
\end{equation}
Since the boundary of the GVT MOND window is given by \eqref{10.22}, the eq. \eqref{10.29BBC} holds true in whole of  the GVT MOND window provided that  
\begin{equation}
\label{10.30}
 a_0 \sqrt{\frac{1}{256 \kappa^4} + 0.65} < 3.25 a_0 ~~\to~~  0.14< |\kappa|.
\end{equation} 
  Notice that when $|\kappa|< 0.14$ then \eqref{1028BBc} is not  valid in a shell adjacent to the boundary of the MOND window.  The physical GEM follows from \eqref{BBPhys7.9}, \eqref{10.23BBc}, \eqref{1024BBc} and \eqref{1028BBc}:
\begin{subequations}
\label{10.31final}
\begin{eqnarray}
\nabla \Phi_{Phy} = \nabla (\Phi_{EH} +  B_0 + \tilde{B}_0) = 4  (0.39 - \kappa) \nabla \Phi_{EH}\,, \\
\nabla \times \vec{A}_{Phy} =\nabla (A_{EH} +  \vec{B} + \vec{\tilde{B}}) =  (1.56- \kappa) \nabla \times \vec{A}_{EH} \,,
\end{eqnarray}
\end{subequations}
We next look at part of the GVT MOND window that holds 
\begin{equation}
\label{10.33}
 |\nabla B_0|^2> |\nabla \times \vec{B}|^2\,.
\end{equation}
wherein \eqref{1026BBc} can be approximated to:
\begin{subequations}
\label{10.34BBd}
\begin{eqnarray}
\frac{ |\nabla {B}_0|}{a_0} \nabla B_0  & = & -\alpha ( 2 z \hat{z} - \rho \hat{\rho}) + \nabla\times \vec{\tilde{h}}\,, \\
\frac{ |\nabla B_0|}{a_0} \nabla\times  \vec{B} & = & 3.22 a_0  \hat{x} + \nabla \tilde{h}\,,\\
\nabla \times \frac{-\alpha ( 2 z \hat{z} - \rho \hat{\rho}) + \nabla\times \vec{\tilde{h}}}{|\nabla B_0|}&=&0\,,\\
\nabla.\frac{3.22 a_0  \hat{x} + \nabla \tilde{h}}{ |\nabla B_0|}&=&0\,. 
\end{eqnarray}
\end{subequations}
Eq. \eqref{10.34BBd} represents a set of second order partial differential equations. It can be analytically solved around $z\approx 0$ or $\rho \approx 0$ where it holds \eqref{1028BBcc}. The solution around $z\approx 0$ reads
\begin{subequations}
\label{10.35}
\begin{eqnarray}
\nabla B_0  & = & +\sqrt{\alpha a_0 \rho }\, \hat{\rho} +O(z)= \sqrt{\frac{a_0}{\alpha \rho}} \nabla \Phi_{EH} + O(z)\,, \\
\nabla\times  \vec{B} & = & 3.22 (\frac{a_0}{\alpha \rho })^{\frac{1}{2}}a_0  \hat{x} +O(z)= \sqrt{\frac{a_0}{\alpha \rho}} \nabla \times A_{EH} + O(z)\,,
\end{eqnarray}
\end{subequations}
while the solution around  $\rho\approx 0$  follows
\begin{subequations}
\label{10.36}
\begin{eqnarray}
\nabla B_0  & = & -\text{Sign}(z)\sqrt{2 \alpha a_0 z }\, \hat{z} +O(\rho)= \sqrt{\frac{a_0}{2 \alpha z}} \nabla \Phi_{EH} + O(\rho)\,, \\
\nabla\times  \vec{B} & = & 3.22 (\frac{a_0}{2 \alpha z })^{\frac{1}{2}}a_0  \hat{x} +O(\rho) =\sqrt{\frac{a_0}{2 \alpha z}} \nabla \times A_{EH} + O(\rho)\,.
\end{eqnarray}
\end{subequations}
Eq. \eqref{10.35} and \eqref{10.36} are respectively valid in
\begin{subequations}
\label{10.37}
\begin{eqnarray}
3.22  a_0 & <\alpha \rho <& \sqrt{\frac{1}{256 \kappa^4} + 0.65} \,a_0 \,,\\
3.22  a_0 & <2 \alpha z <& \sqrt{\frac{1}{256 \kappa^4} + 0.65} \,a_0 \,,
\end{eqnarray} 
\end{subequations}
wherein the upper bound is the boundary of the MOND regime and the lower bound is \eqref{10.33} written for \eqref{10.35} and \eqref{10.36}. This means that  \eqref{10.35} and \eqref{10.36} are valid solutions provided that  
\begin{equation}
  |\kappa| < 0.14,
\end{equation} 
which is complementary to \eqref{10.30}. Eq.  \eqref{10.35} and \eqref{10.36} describe the behavior of $B$ in a shell adjacent to the boundary of the MOND window that eq. \eqref{1028BBc}  is not valid in. They present an enhancement for the gravitomagnetic and gravitoelectric field strengths. However due to the lower bound in \eqref{10.37}, the enhancement is bounded and is not akin to  that  of naive extension of  MOND to the gravitomagnetic force:  eq. \eqref{A.10} for $\rho\to 0$ or $z\to 0$. 
 
 The physical GEM following  from \eqref{BBPhys7.9},  \eqref{10.23BBc}, \eqref{1024BBc} for \eqref{10.35} reads:
 \begin{subequations}
 \label{10.38}
 \begin{eqnarray}
 \nabla \Phi_{Phy} = \left(1- 4 \kappa +  \sqrt{\frac{a_0}{2 \alpha z}}\right) \nabla \Phi_{EH}+ O(\rho)\,, \\
\nabla \times \vec{A}_{Phy} =\left(1-  \kappa +  \sqrt{\frac{a_0}{2 \alpha z}}\right) \nabla \times \vec{A}_{EH} + O(\rho)\,.
 \end{eqnarray}
\end{subequations}
While \eqref{10.36} yields:
 \begin{subequations}
 \label{10.39}
 \begin{eqnarray}
 \nabla \Phi_{Phy} = \left(1- 4 \kappa +  \sqrt{\frac{a_0}{\alpha \rho}}\right) \nabla \Phi_{EH}+ O(z)\,, \\
\nabla \times \vec{A}_{Phy} =\left(1-  \kappa +  \sqrt{\frac{a_0}{ \alpha \rho}}\right) \nabla \times \vec{A}_{EH} + O(z)\,.
 \end{eqnarray}
\end{subequations} 
Eq. \eqref{10.38} and \eqref{10.39} respectively describe the GEM  around $z\approx 0$ and $\rho \approx 0$ for the GVT  MOND regime that holds \eqref{10.33}. The physical GEM in other points will be identified after solving \eqref{10.28BBc} and choosing the boundary conditions on $\vec{\tilde{h}}$ and $\tilde{h}$  such that the general solution reduces to  \eqref{10.38} and \eqref{10.39} respectively  for $z\approx 0$ and $\rho \approx 0$.

The accurate tracking of a probe passing  through the MOND windows is the simplest way to test the physics of the Solar MOND windows. For $k\ge 0.14$, a probe that passes through the Sun-Jupiter GVT MOND window experiences the following anomalous acceleration:
\begin{equation}
\label{10.40}
\vec{a}_{\text{anomaly}} \,=\, -(0.56- 4 \kappa) \nabla \Phi_{EH}+ (0.56 - \kappa) \frac{\vec{v}}{c}\times \nabla \times \vec{A}_{EH}\,,
\end{equation}
where $\vec{v}$ is the velocity of the probe with respect to the Sun and we have utilized \eqref{appendinxneeds} and \eqref{10.31final}.  For $k< 0.14$, a probe moving in $z=0$ or $\rho=0$ experiences the following anomalous acceleration in the regime of \eqref{10.37}
\begin{equation}
\vec{\tilde{a}}_{\text{anomaly}} \,=\, -\left(-4 \kappa+\sqrt{\frac{a_0}{|\nabla \Phi_{EH}|}}\right) \nabla \Phi_{EH}+\left(-\kappa+\sqrt{\frac{a_0}{|\nabla \Phi_{EH}|}}\right) \frac{v}{c}\times \nabla \times \vec{A}_{EH}\,,
\end{equation}
while experiences the anomalous acceleration given by \eqref{10.40} in the rest of the GVT MOND window. We observe that for the peculiar value of $k=0.14$, a slow moving probe ($\frac{|\vec{v}|}{c}\ll 1$) will not experience an anomalous acceleration inside the Sun-Jupiter MOND window. This peculiar value of $k$ is not universal and depends on the details of the considered MOND window. In order to refute the GVT theory by the accurate tracking of a probe that passes through the MOND windows,  therefore, we must either
\begin{itemize}
\item increase the precision such that the anomalous acceleration in the gravitomagnetic force be observed,
\item or to track probes in different MOND windows.
\end{itemize}
 Observing an anomaly in a single MOND window, however, refutes the Einstein-Hilbert theory  and favors the GVT, TeVeS or the Moffat's theory. The GVT theory, so far, is the only generally covariant theory that also predicts an anomaly in the gravitomagnetic field inside the MOND window.

 \section{Conclusion and outlook}
\label{section11}
We have introduced the Gauge Vector Tensor theory: a generally covariant theory of gravity   composed of a pseudo Riemannian metric and two $U(1)$ gauge connections that reproduces  MOND in  the limit of very weak gravitational fields  while remains consistent with the Einstein-Hilbert gravity in the limit of strong and Newtonian gravitational fields. The nonlinearity introduced by the GVT theory to reproduce the MOND behavior resides only inside the MOND regime and it does not propagates to the strong regime of gravity. This is a clear advantage of the GVT theory over the Bekenstein's Tensor-Vector-Scaler theory \cite{Bekenstein:2004ne}. We have been motivated to introduce the GVT theory after uplifting  the GravitoElectroMagnetism approximation to gravity to the Milgrom's MOND theory \cite{MOND}.    

  We have illustrated that  the  gravitomagnetic force at the edge of a galaxy can  be in accord with either GVT  or $\Lambda$CDM but not both.   We also have studied the physics of the GVT theory around  the  gravitational saddle point of the Sun and Jupiter system. We have noticed that the conclusive  refusal of  the GVT theory demands measuring    either  both of the gravitoelectric and gravitomagnetic fields inside the Sun-Jupiter  MOND window, or the gravitoelectric field inside two different solar GVT MOND windows. The GVT theory, however, can be favored by observing an anomaly in the gravitoelectric field inside a single MOND window.
 
 We also need to study the cosmology and the gravitational lensing of the GVT theory.   Let it be hasten that, as shown in section \ref{section6},  the GVT theory is an extension of the Moffat's Scalar-Tensor-Vector theory \cite{Moffat:2005si}. We, therefore, envisage that it inherits most of the merits of the Moffat's theory  in describing the gravitational lensing and cosmology.   We, however, accomplish this study elsewhere.   
 \section*{Acknowledgements}
 This work was supported by the Institute for Research in Fundamental Sciences.

 \appendix
 \section{Naive extension of MOND to  the GravitoMagnetic Force }
 Modified Newtonian Dynamics (MOND)  provides an alternative approach to the missing mass problem in galaxies. It assumes that the newtonian dynamics is governed by 
\begin{equation}
\vec{F} = m f(\frac{|a|}{a_0}) \vec{a}
\end{equation}
where $F$ is the force exerted on the center of the mass of the object, $a$ is the acceleration of the object with respect to the cosmological frame wherein the CMB is uniform, and  $a_0$ is given in \eqref{valuea0mo}.
MOND coincides to the Newtonian dynamics  in large accelerations:
 \begin{equation}
 \lim_{x\to \infty} f(x)=1\,.
 \end{equation}
 Note that $x\ll 1$ is called the Newtonian regime of the MOND theory.
 To account for the missing mass problem, it is required that 
 \begin{equation}
  f(x)=x\, \qquad\mbox{for} ~x\le 1\,.
 \end{equation}
 Note that  $x\le 1$ is called the MOND regime. 

The gravitational force extorted on a slow moving particle (the test particle) of mass $m$ and velocity $\vec{v}$ in the GravitoElectroMagnetism  approximation  to gravity follows from \eqref{appendinxneeds}
\begin{equation}
F =  m (-\nabla \phi +  \vec{v}\times \nabla\times A)   
\end{equation}
where $\phi$ is the Newtonian gravitational field (the gravitoelectric field) and $\nabla\times A$ is the gravitomagnetic field strength. The gravitoelectromagnetic fields of  a spherical static mass distribution read
\begin{subequations}
\label{APhi}
\begin{eqnarray}
\phi(r) & = & - \frac{G M}{r}\,, \\
A(r) & = & -\frac{2 G}{c^2 r^3} \vec{J} \times \vec{r}\,,
\end{eqnarray}
\end{subequations}
where $r$ is the distance from the center of the mass distribution (the source), $M$ is its total mass and $J$  represents the total angular momentum of the source.

The GEM approximation of the Newtonian regimes of the MOND paradigm coincide to that of the the Einstein-Hilbert gravity. The story, however, changes in the MOND regime of the theory.  The MOND regime holds
\begin{equation}\label{MondGEM}
-\nabla \phi +  \vec{v}\times \nabla\times A = \frac{|a|}{a_0} a\,.
\end{equation}
which is a non-linear second order differential equation for the position of the test particle. Eq. \eqref{MondGEM} results 
\begin{equation}
\frac{|a|^4}{a_0^2} = |-\nabla \phi +  \vec{v}\times \nabla\times A|^2
\end{equation}
using which in  \eqref{MondGEM}  returns
\begin{equation}
\label{aMGEMexact}
a= \sqrt{\frac{a_0}{|-\nabla \phi +  \vec{v}\times \nabla\times A|}}(-\nabla \phi +  \vec{v}\times \nabla\times A) \,,
\end{equation}
Around galaxies the gravitoelectric force is much larger than the gravitomagnetic one. We therefore can taylor expand \eqref{aMGEMexact} in term of $v$ and obtain:
\begin{equation}\label{1.11mond}
a= \sqrt{\frac{a_0}{|\nabla \phi|}}\left(-\nabla \phi +  \vec{v}\times \nabla\times A + \nabla\phi \frac{\nabla \phi.\vec{v}\times(\nabla\times A)}{|\nabla \phi|^2}\right) +O(v^2)\,.
\end{equation}
The first two terms in the r.h.s of \eqref{1.11mond} can be interpreted as the gravitoelectric and gravitomagnetic force in the the MOND regime:
\begin{subequations}
\label{A.10}
\begin{eqnarray}
E_{\mbox{\tiny MOND}} &=& \sqrt{\frac{a_0}{|\nabla \phi|}}\nabla \phi\,,\\
B_{\mbox{\tiny MOND}} &=&  \sqrt{\frac{a_0}{|\nabla \phi|}}  \nabla\times A\,. 
\end{eqnarray}
\end{subequations}
Since the MOND regime holds $|\nabla \phi|<a_0$ then we observe an enhancement in the gravitoelectric and gravitomagnetic field strength. The enhancement factor is $\sqrt{\frac{a_0}{|\nabla \phi|}}$. The last term in the r.h.s of  \eqref{1.11mond} is a new kind of gravitational force acting on the test particle. This new force  can be expressed through 
\begin{equation}
F = m  \nabla \phi \frac{\vec{v}.\nabla \phi \times (\nabla \times A)}{|\nabla \phi|^2} \sqrt{\frac{a_0}{|\nabla \phi|}}\,,
\end{equation}
Two understand a possible meaning of this force let us consider the gravitoelectric and gravitomagnetic field strength of  a spherical static mass distribution given in \eqref{6162B2}.
The new force then simplifies to 
\begin{equation}\label{1.17}
F = \sqrt{\frac{4 G a_0}{M c^4}}\, \frac{\vec{v}.\vec{r}\times J }{r^3} \hat{r}
\end{equation}
Note that $J$ is the total angular momentum of the mass distribution and it is proportional to the total mass. Therefore the limit of $M\to 0$ in the eq. \eqref{1.17} exists. This force is in the direction of the gravitoelectric force and is less than it. So it would not significantly change the physics. We, however,  take the position  that this new force is an artifact of naively applying the MOND to the  gravitomagnetic force.

\providecommand{\href}[2]{#2}\begingroup\raggedright

\end{document}